\documentclass[preprintnumbers,amsmath,amssymb,floatfix,12pt,prd,superscriptaddress,nofootinbib]{revtex4}
\usepackage{graphicx}
\usepackage{epsfig}
\usepackage{bm}
\usepackage{amsfonts}
\usepackage{subfigure}

\begin{document}

\title{Generalized second law of thermodynamics in scalar-tensor gravity}

\author{A. Abdolmaleki}
\email{AAbdolmaleki@uok.ac.ir}\affiliation{Research Institute for
Astronomy and Astrophysics of Maragha (RIAAM), P.O. Box 55134-441,
Maragha, Iran}

\author{T. Najafi}
\email{t.najafi90@gmail.com} \affiliation{Department of Physics,
University of Kurdistan, Pasdaran St., P.O. Box 66177-15175,
Sanandaj, Iran}

\author{K. Karami}
\email{KKarami@uok.ac.ir}\affiliation{Department of Physics,
University of Kurdistan, Pasdaran St., P.O. Box 66177-15175,
Sanandaj, Iran}

\date{\today}

\begin{abstract}
\vspace*{0.5cm} \centerline{\bf Abstract} \vspace*{0.5cm} Within the
context of scalar-tensor gravity, we explore the generalized second
law (GSL) of gravitational thermodynamics. We extend the action of
ordinary scalar-tensor gravity theory to the case in which there is
a non-minimal coupling between the scalar field and the matter field
(as chameleon field). Then, we derive the field equations governing
the gravity and the scalar field. For a FRW universe filled only
with ordinary matter, we obtain the modified Friedmann equations as
well as the evolution equation of the scalar field. Furthermore, we
assume the boundary of the universe to be enclosed by the dynamical
apparent horizon which is in thermal equilibrium with the Hawking
temperature. We obtain a general expression for the GSL of
thermodynamics in the scalar-tensor gravity model. For some viable
scalar-tensor models, we first obtain the evolutionary behaviors of
the matter density, the scale factor, the Hubble parameter, the
scalar field, the deceleration parameter as well as the effective
equation of state (EoS) parameter. We conclude that in most of the
models, the deceleration parameter approaches a de Sitter regime at
late times, as expected. Also the effective EoS parameter acts like
the $\Lambda$CDM model at late times. Finally, we examine the
validity of the GSL for the selected
models.
\end{abstract}



\maketitle


\section{Introduction}
During the last decade, observational cosmology has entered an era
of unprecedented precision. Measurements of the cosmic microwave
background (CMB) (\cite{Komatsu,2}), the Hubble constant ($H_0$)
\cite{3}, the luminosity and distance at high redshift with
supernovae Ia \cite{4}, and baryon acoustic oscillations (BAO)
surveys \cite{5}, suggest that our Universe is currently undergoing
a phase of accelerated expansion. The proposals that have been put
forth to explain these interesting discoveries can basically be
classified into two categories. One is to assume the cosmic speed-up
might be caused within general relativity (GR) by a mysterious
cosmic fluid with negative pressure, which is usually called dark
energy (DE). However, the nature of DE is still unknown and the
problem of DE is one of the hardest and unresolved problems in
modern theoretical physics (see \cite{Padmanabhan,Sahni1} and
references therein).

Alternatively, the acceleration could be due to purely gravitational
effects, named modified gravity, i.e., one may consider modifying
the current gravitational theory to produce an effective DE. One
such modification is referred to as $f(R)$-gravity, in which the
Einstein-Hilbert action in GR is generalized from the Ricci scalar
$R$ to an arbitrary function of the Ricci scalar (for a good review
see \cite{Sotiriou} and references therein). There are also some
other classes of modified gravities containing $f(\mathcal{G})$
\cite{fG}, $f(R,\mathcal{G})$ \cite{fRG} and $f(T)$ \cite{fT} which
are considered as gravitational alternatives for DE. Here,
${\mathcal
G}=R_{\mu\nu\rho\sigma}R^{\mu\nu\rho\sigma}-4R_{\mu\nu}R^{\mu\nu}+R^2$
is the Gauss-Bonnet invariant term. Also $R_{\mu\nu\rho\sigma}$ and
$R_{\mu\nu}$ are the Riemann and Ricci tensors, respectively, and
$T$ is the torsion scalar. The modified gravity can unify the
early-time inflation with late-time acceleration without resorting
to the DE \cite{Sotiriou}. Moreover, modified gravity may serve as
dark matter (DM) \cite{Sobouti}.

In the context of modified gravity, there is also a large class of
models called scalar-tensor theories \cite{Faraoni,CapReview}, which
take into account the effects due to the non-minimal coupling term
$F(\phi)R$ between a scalar field $\phi$ and a Ricci scalar
curvature. In scalar-tensor theories, if the evolution of matter
perturbations $\delta_{\rm m}=\delta\rho_{\rm m}/\rho_{\rm m}$ is
known observationally, together with the Hubble parameter $H(z)$,
one can even determine the function $F(\phi)$ together with the
potential $V(\phi)$ of the scalar field \cite{Nesseris}.
Scalar-tensor theories also contains a class of models called
chameleon gravity \cite{Khoury,Khoury1,Saaidi} in which there is a
non-minimal coupling between the scalar field and the matter field.
Historically, one of the first scalar-tensor theories is the
Brans-Dicke theory of gravity which has been motivated from Mach's
principle. This is achieved in Brans-Dicke theory by making the
effective gravitational coupling strength $G_{\rm eff}\sim
\phi^{-1}$ depend on the space-time position and being governed by
distant matter sources. Modern interest in Brans-Dicke and
scalar-tensor theories is motivated by the fact that they are
obtained as low-energy limits of string theories. It was shown that
metric and Palatini (but not metric-affine) modified gravities can
be reduced to scalar-tensor theories \cite{Faraoni2}.

Thermodynamics of the accelerating universe driven by the DE or dark
gravity (due to the modified gravity effect) is one of interesting
issues in modern cosmology. In the context of black hole
thermodynamics, Jacobson using the first law of thermodynamics on
the local Rindler horizons and assuming the Bekenstein-Hawking
entropy-area relation $S_{\rm BH}=A/(4G)$, where $A$ is the area of
the horizon and $G$ is Newton's constant, was able to derive the
Einstein equations \cite{Jacobson}. The study on the connection
between gravity and thermodynamics has been extended to cosmological
context. It was pointed out that the Friedmann equation in the
Einstein gravity can be obtained using the first law of
thermodynamics (Clausius relation) $-{\rm d}E=T_A{\rm d} S_A$ on the
apparent horizon $\tilde{r}_{\rm A}$ with the Hawking temperature
$T_{\rm A}=1/(2\pi \tilde{r}_{\rm A})$ and Bekenstein-Hawking
entropy $S_A=\frac{A}{4G}$ \cite{Cai05}. The relation between
gravity and thermodynamics has been further disclosed in extended
gravitational theories, including the $f(\mathcal{G})$ theory
\cite{Cai05}, scalar-tensor gravity and $f(R)$-gravity
\cite{Akbar12}, Lovelock theory \cite{Akbar} and braneworld
scenarios (such as DGP, RSI and RSII) \cite{Sheykhi1}.

Note that the entropy-area relation $S_A=\frac{A}{4G}$ familiar from
GR is still valid in the other modified gravity theories provided
that Newton's constant $G$ is replaced by a suitable effective
gravitational coupling strength $G_{\rm eff}$. For instance, the
effective Newton's constant in $f(R)$-gravity and $f(T)$-gravity are
given by $G_{\rm eff}=G/f'(R)$ \cite{Wald} and $G_{\rm eff}=G/f'(T)$
\cite{Miao}, respectively, where prime denotes a derivative with
respect to the Ricci $R$ and torsion $T$ scalars. In scalar-tensor
gravity, the geometric entropy is also given by
$S_A=\frac{A}{4G_{\rm eff}}$ \cite{Faraoni2,fphi} with $G_{\rm
eff}=G/F(\phi)$ \cite{Faraoni2,fphi}.

In addition to the first law of thermodynamics, the generalized
second law (GSL) of gravitational thermodynamics, which states that
entropy of the fluid inside the horizon plus the geometric entropy
do not decrease with time, has been studied extensively in the
literature \cite{Izquierdo1}-\cite{Geng}. The GSL of thermodynamics
like the first law is a universal principle governing the Universe.
Here, our aim is to investigate the GSL of thermodynamics in the
framework of scalar-tensor gravity. As one of the most important
theoretical touch stones to examine whether scalar-tensor gravity
can be an alternative gravitational theory to GR, we explore the GSL
of thermodynamics in scalar-tensor gravity, and derive the condition
for the GSL to be satisfied. The paper is organized as follows. In
section \ref{STg} we investigate the scalar-tensor gravity and
extend it to the case in which there is a non-minimal coupling
between the scalar field and the matter field (as chameleon field).
In section \ref{GSL-STg}, we explore the GSL of thermodynamics on
the dynamical apparent horizon of a Friedmann-Robertson-Walker (FRW)
universe filled with the ordinary matter which is in thermal
equilibrium with the Hawking temperature. In sections
\ref{Model_I}-\ref{Model_V}, we examine the validity of the GSL for
some viable scalar-tensor gravity models containing Brans-Dicke
gravity, Brans-Dicke gravity with a self interacting potential,
chameleon gravity, chameleonic generalized Brans-Dicke gravity and
chameleonic Brans-Dicke gravity with a self interacting potential.
Section \ref{Conclusions} is devoted to conclusions.
\section{Scalar-tensor gravity}\label{STg}
In the Jordan frame, general action of the scalar-tensor gravity can
be written as
\begin{equation}
I= \int {\rm d}^{4}x\sqrt{-g}\left[\frac{1}{2k^{2}}\Big(F(\phi)R -
Z(\phi)g^{\mu\nu}\phi_{,\mu}\phi_{,\nu}-2U(\phi)\Big)+E(\phi)L_{\rm
m}\right],\label{action}
\end{equation}
where $k^2=8\pi G$. Also $g$, $R$, $\phi$ and $L_{\rm m}$ are the
determinant of metric $g_{\mu\nu}$, the Ricci scalar curvature, the
scalar field and the matter Lagrangian, respectively. Also
$F(\phi)$, $Z(\phi)$ and $E(\phi)$ are arbitrary dimensionless
functions, and $U(\phi)$ is the scalar field potential. Note that in
action (\ref{action}), the terms $F(\phi)R$ and $E(\phi)L_{\rm m}$
show that the scalar field $\phi$ is nonminimally coupled to the
scalar curvature (as Brans-Dicke field with $F(\phi)=\phi$) and the
matter Lagrangian (as chameleon field), respectively. In the absence
of chameleon field, i.e. $E(\phi)=1$, Eq. (\ref{action}) reduces to
the ordinary action of the scalar-tensor gravity theory
\cite{Primo}.

Taking variations of the action (\ref{action}) with respect to
$g_{\mu\nu}$ and $\phi$ lead to the corresponding field equations in
scalar-tensor gravity as
\begin{eqnarray}\label{field}
F(\phi)G_{\mu\nu}=k^{2}T^{\rm m}_{\mu\nu}E(\phi)+Z(\phi)
\Big[\partial_{\mu}\phi\partial_{\nu}\phi-\frac{1}{2}g_{\mu\nu}\big(\partial_{\alpha}\phi\big)^{2}\Big]
\nonumber\\+\nabla_{\mu}\partial_{\nu}F(\phi)-g_{\mu\nu}\Box
F(\phi)-g_{\mu\nu}U(\phi),
\end{eqnarray}
\begin{equation}\label{phiEq}
2Z(\phi)\Box\phi= 2U_{,\phi}-F_{,\phi}R-
Z_{,\phi}(\partial_{\alpha}\phi)^{2}-
\frac{k^2}{2}g^{\mu\nu}E_{,\phi}T^{\rm m}_{\mu\nu},
\end{equation}
where $ G_{\rm\mu\nu}=R_{\mu\nu}-\frac{1}{2}Rg_{\mu\nu}$ is the
Einstein tensor, $R_{\mu\nu}$ is the Ricci tensor and $T^{\rm
m}_{\mu\nu}$ is the energy-momentum tensor of the matter fields.
Also $\nabla_{\mu}$ is the covariant derivative associated with
$g_{\mu\nu}$ and the subscript $\phi$ denotes a derivative with
respect to the scalar field $\phi$ (i.e. $F_{,\phi} = {\rm d}F/{\rm
d}\phi$). We assume that $T^{\rm m}_{\mu\nu}$ has the form of the
energy-momentum tensor of a perfect fluid
\begin{equation}\label{TEq}
T^{\rm m}_{\mu\nu}=p_{\rm m}g_{\mu\nu}+ (p_{\rm m}+\rho_{\rm
m})U_{\mu}U_{\nu}.
\end{equation}
Now we consider a spatially non-flat universe described by the FRW
metric
\begin{equation}\label{FRW 1}
{\rm d}s^2 = -{\rm d}t^2+a^2(t)\left(\frac{{\rm d}r^2}{1-Kr^2} +
{r}^2{\rm d}\Omega^2\right),
\end{equation}
where $K=0,1,-1$ represent a flat, closed and open universe,
respectively. Substituting the FRW metric (\ref{FRW 1}) into the
field equations (\ref{field}) yields the Friedmann equations in
scalar-tensor gravity as
\begin{equation}\label{Fr 1}
3F(\phi)\left(H^{2}+\frac{K}{a^{2}}\right)=k^{2}\rho_{\rm m}
E(\phi)+\frac{Z(\phi)}{2}~\dot{\phi}^{2}-3H\dot{F}+ U(\phi),
\end{equation}
\begin{equation}\label{Fr 2}
-2F(\phi)\left(\dot{H}-\frac{K}{a^{2}}\right)=k^{2}(\rho_{\rm
m}+p_{\rm m})E(\phi)+Z(\phi)\dot{\phi}^{2}+ \ddot{F}- H\dot{F}.
\end{equation}
Also Eq. (\ref{phiEq}) for the FRW metric (\ref{FRW 1}) gives the
equation governing the evolution of the scalar field as
\begin{equation}\label{evolu}
2Z(\phi)\left(\ddot{\phi}+3H\dot{\phi}\right)= RF_{,\phi}-
Z_{,\phi}\dot{\phi}^{2}-2U_{,\phi}-\frac{k^2}{2}E_{,\phi}(\rho_{\rm
m}-3p_{\rm m}),
\end{equation}
where
\begin{equation}
R=6\left(\dot{H}+2H^{2}+ \frac{K}{a^{2}}\right),
\end{equation}
and $H=\dot{a}/a$ is the Hubble parameter. Here the dot denotes a
derivative with respect to cosmic time $t$. Note that in the absence
of chameleon field, i.e. $E(\phi)=1$, Eqs. (\ref{Fr 1}), (\ref{Fr
2}) and (\ref{evolu}) are same as those obtained for the ordinary
scalar-tensor gravity \cite{Primo}.

The Friedmann equations (\ref{Fr 1}) and (\ref{Fr 2}) can be
rewritten in the standard form as
\begin{equation}
H^2+\frac{K}{a^2}=\frac{k^2}{3}\rho_{\rm eff},\label{Fr}
\end{equation}
\begin{equation}
\dot{H}-\frac{K}{a^{2}}=-\frac{k^2}{2}(\rho_{\rm eff}+p_{\rm
eff}),\label{Hdot}
\end{equation}
where $\rho_{\rm eff}$ and $p_{\rm eff}$ are the effective (total)
energy density and pressure defined as
\begin{eqnarray}
&&\rho_{\rm eff}=\frac{1}{F(\phi)}\left(\rho_{\rm
m}E(\phi)+\frac{\rho_\phi}{k^{2}}\right),\\
&&p_{\rm eff}=\frac{1}{F(\phi)}\left(p_{\rm
m}E(\phi)+\frac{p_{\phi}}{k^{2}}\right).
\end{eqnarray}
Here $\rho_\phi$ and $p_\phi$ are the energy density and pressure
due to the scalar field contribution defined as
\begin{equation}\label{roR}
\rho_{\phi}= \frac{Z(\phi)}{2}~\dot{\phi}^{2}-3H\dot{F}+ U(\phi),
\end{equation}
\begin{equation}\label{pR}
p_{\phi}=\frac{Z(\phi)}{2}~\dot{\phi}^{2}+ \ddot{F}+ 2H\dot{F}-
U{\rm(\phi)}.
\end{equation}
Note that the scalar field contributions $\rho_{\phi}$ and
$p_{\phi}$ in scalar-tensor gravity can justify the observed
acceleration of the universe without resorting to the DE. For a
special case $F(\phi)=E(\phi)=1$, from Eqs. (\ref{roR}) and
(\ref{pR}) we have $\rho_{ \phi}= \frac{Z(\phi)}{2}~\dot{ \phi}^{2}+
U(\phi)$ and $p_{\phi}=\frac{Z(\phi)}{2}~\dot{\phi}^{2}- U(\phi)$,
then Eqs. (\ref{Fr}) and (\ref{Hdot}) transform to the usual
Friedmann equations in the Einstein gravity.

The energy conservation laws in scalar-tensor gravity can be
obtained as
\begin{equation}
\dot{\rho}_{\rm m}+3H\left( \rho_{\rm m}+p_{\rm m}\right)
=-\frac{3}{4}(\rho_{\rm m}+p_{\rm
m})\frac{\dot{E}(\phi)}{E(\phi)},\label{ec}
\end{equation}
\begin{equation} \label{tot con}
\dot{\rho}_{\rm eff}+3H\left( \rho_{\rm eff}+p_{\rm eff}\right) =0.
\end{equation}
Also $\rho_\phi$ and $p_\phi$ satisfy the following energy equation
\begin{equation}\label{R con}
\dot{\rho}_{\phi}+3H\left( \rho_{\phi}+p_{\phi}\right)
=k^2\left[\rho_{\rm
eff}\dot{F}(\phi)-\frac{1}{4}\dot{E}(\phi)(\rho_{\rm m}-3p_{\rm
m})\right].
\end{equation}
Note that the set of equations containing the Friedmann equations
(\ref{Fr 1}) and (\ref{Fr 2}), the evolution equation of the scalar
field (\ref{evolu}) and the continuity equation governing the matter
field (\ref{ec}) are not independent of each other. Taking the time
derivative of Eq. (\ref{Fr 1}) and using Eqs. (\ref{evolu}) and
(\ref{ec}), one can get the second Friedmann equation (\ref{Fr 2}).
In the next sections, we take the set of Eqs. (\ref{Fr 2}),
(\ref{evolu}) and (\ref{ec}) which can uniquely determine the
dynamics of the Universe.

\section{GSL in scalar-tensor gravity}\label{GSL-STg}

Here in the context of scalar-tensor gravity theory, we explore the
GSL of gravitational thermodynamics on the dynamical apparent
horizon of a FRW universe filled only with ordinary matter which is
in thermal equilibrium with the Hawking temperature. The GSL states
that the sum of entropy of fluid filling the universe along with the
entropy of the cosmological horizon must be increasing (or
non-decreasing) function of time \cite{Cai05}.

For a spatially non-flat FRW universe, the dynamical apparent
horizon takes the form \cite{Poisson}
\begin{equation}\label{ra}
\tilde{r}_{\rm A}=\left( H^{2}+\frac{K}{a^{2}}\right)^{-1/2},
\end{equation}
which in the case of flat universe ($K = 0$), it reduces to the
Hubble horizon, i.e. $\tilde{r}_{\rm A}=H^{-1}$. On the apparent
horizon, the associated Hawking temperature is defined as
\cite{Cai05}
\begin{equation}
T_{\rm A}=\frac{1}{2\pi \tilde{r}_{\rm
A}}\left(1-\frac{\dot{\tilde{r}}_{\rm A}}{2H\tilde{r}_{\rm
A}}\right),\label{ta}
\end{equation}
where the condition $\frac{\dot{\tilde{r}}_{A}}{2H\tilde{r}_{A}}<1$
is necessary due to having a positive temperature. Cai et al.
\cite{Cai09} using the tunneling approach, proved that there is
indeed a Hawking radiation with temperature (\ref{ta}), for a
locally defined apparent horizon of a FRW universe with any spatial
curvature.

The entropy of the matter inside the horizon satisfies the Gibbs
equation \cite{Izquierdo1}
\begin{equation}\label{gi}
T_{\rm A}{\rm d}S_{\rm m}={\rm d}E_{\rm m}+p_{\rm m}{\rm d}{V},
\end{equation}
where $E_{\rm m}=\rho_{\rm m}{V}$ and $V=\frac{4\pi}{3}
\tilde{r}_{\rm A}^{3}$ is the volume of the dynamical apparent
horizon $\tilde{r}_{\rm A}$ containing the matter. Here, we have
assumed the local equilibrium hypothesis to hold
\cite{Pavon2,KG2010}.

Taking the time derivative of Eq. (\ref{gi}) and using (\ref{ec})
gets the equation governing the evolution of the matter entropy as
\begin{equation}\label{S mat f}
T_{\rm A}\dot{S_{\rm m}}=4\pi\tilde{r}_{\rm A}^{3}(\rho_{\rm
m}+p_{\rm m})\left(\frac{\dot{\tilde{r}}_{\rm A}}{\tilde{r}_{\rm
A}}-H- \frac{1}{4}\frac{\dot{E}}{E}\right).
\end{equation}
Using Eq. (\ref{Fr 2}), this can be rewritten as
\begin{equation}
T_{\rm A}\dot{S}_{\rm m}=\frac{-\tilde{r}_{\rm A}^{2}}{2G~E{\rm
(\phi)}}\left(\dot{\tilde{r}}_{\rm A}-H\tilde{r}_{\rm A}-
\frac{1}{4}\frac{\dot{E}}{E}~\tilde{r}_{\rm A}\right)
\left[\left(2\dot{H}-\frac{2K}{a^{2}}-H\frac{\rm d}{{\rm d}
t}+\frac{{\rm d}^{2}}{{\rm d} t^{2}}\right)F{\rm(\phi)}+~Z{\rm
(\phi)}\dot{\phi}^{2}\right].\label{sma}
\end{equation}
The geometric entropy in the scalar-tensor gravity is given by
\cite{fphi}
\begin{equation}\label{S hor}
S_{\rm A}= \frac{AF{\rm (\phi)}}{4G},
\end{equation}
where ${\rm A}=4\pi\tilde{r}_{\rm A}^{2}$ is the area of the
apparent horizon. Taking the time derivative of Eq. (\ref{S hor})
and using (\ref{ta}) yields the evolution of the horizon entropy as
\begin{equation}\label{saa}
T_{\rm A}\dot{S}_{\rm A}=\frac{\tilde{r}_{\rm
A}}{4GH}\left(2H-\frac{\dot{\tilde{r}}_{\rm A}}{\tilde{r}_{\rm
A}}\right)\left(\frac{2\dot{\tilde{r}}_{\rm A}}{\tilde{r}_{\rm
A}}+\frac{\rm d}{{\rm d} t}\right)F(\phi).
\end{equation}
Now according to the GSL of gravitational thermodynamics, we can
consider the entropy of the universe as the sum of the entropy of
the matter inside the horizon, and the horizon entropy. Adding Eqs.
(\ref{sma}) and (\ref{saa}) and using the auxiliary relation
\begin{eqnarray}
&&\dot{\tilde{r}}_{\rm A}=H\tilde{r}_{\rm
A}^{3}\left(\frac{K}{a^{2}}-\dot{H}\right),\label{dra}
\end{eqnarray}
the GSL in scalar-tensor gravity reads
\begin{eqnarray}\label{S Tot}
T_{\rm A}\dot{S}_{\rm
tot}=\frac{1}{4G}\left(H^{2}+\frac{K}{a^{2}}\right)^{-5/2}
\left[{\mathcal J}_1F(\phi)+{\mathcal J}_2\dot{F}(\phi)+{\mathcal
J}_3\Big(Z(\phi)\dot{\phi}^{2}+\ddot{F}(\phi)\Big)\right],
\end{eqnarray}
where
\begin{equation}
{\mathcal J}_1=\left(\frac{K}{a^{2}}-\dot{H}\right)
\left\{2H\left[2H^{2}\left(1-\frac{1}{E}\right)+\dot{H}\left(1-\frac{2}{E}\right)+
\frac{K}{a^{2}}\right]-\frac{\dot{E}}{E^{2}}
\left(H^{2}+\frac{K}{a^{2}}\right)\right\},
\end{equation}
\begin{equation}
{\mathcal
J}_2=\frac{K}{a^{2}}\left(\frac{K}{a^{2}}+\dot{H}+3H^{2}\right)+H^{2}\left[2H^{2}
\left(1-\frac{1}{E}\right)+\dot{H}\left(1-\frac{2}{E}\right)\right]-\frac{H\dot{E}}{2E^{2}}
\left(H^{2}+\frac{K}{a^{2}}\right),
\end{equation}
\begin{equation}
{\mathcal
J}_3=\frac{2H}{E}\Big(\dot{H}+H^{2}\Big)+\frac{\dot{E}}{2E^{2}}
\left(H^{2}+\frac{K}{a^{2}}\right),
\end{equation}
and $S_{\rm tot}=S_{\rm m}+S_{\rm A}$. Equation (\ref{S Tot}) shows
that the validity of the GSL, i.e. $T_{\rm A}\dot{S}_{\rm tot}\geq
0$, depends on the scalar-tensor gravity model. For instance, in the
Einstein gravity, i.e. $F{\rm (\phi)}=E{\rm (\phi)}=1$ and
$Z(\phi)=U(\phi)=0$, the GSL (\ref{S Tot}) yields
\begin{equation}\label{GSL-R}
T_{\rm A}\dot{S}_{\rm tot}=\frac{H}{2G}
\frac{(\dot{H}-\frac{K}{a^{2}})^{2}}{(H^{2}+\frac{K}{a^{2}})^{5/2}}\geq
0,
\end{equation}
which shows that the GSL in Einstein's gravity is always satisfied.

Here following \cite{Radicella2}, we try to rewrite the GSL (\ref{S
Tot}) in terms of the effective equation of state (EoS) parameter
$w_{\rm eff}$ defined as
\begin{equation}\label{weff1}
w_{\rm eff}=\frac{p_{\rm eff}}{\rho_{\rm
eff}}=-1-\frac{2}{3}\left(\frac{\dot{H}-\frac{K}{a^2}}{H^2+\frac{K}{a^2}}\right),
\end{equation}
where we have used Eqs. (\ref{Fr}) and (\ref{Hdot}). From Eqs.
(\ref{ra}), (\ref{dra}) and (\ref{weff1}) one can get
\begin{equation}
\frac{\dot{\tilde{r}}_{\rm A}}{\tilde{r}_{\rm
A}}=\frac{3}{2}H(1+w_{\rm eff}).
\end{equation}
Replacing this into Eqs. (\ref{S mat f}) and (\ref{saa}) yield
\begin{equation}\label{TASdotm}
T_{\rm A}\dot{S}_{\rm m}=\frac{\tilde{r}^3_{\rm A}}{4G}\rho_{\rm
m}(1+w_{\rm m})\left[H(1+3w_{\rm
eff})-\frac{1}{2}\frac{\dot{E}(\phi)}{E(\phi)}\right],
\end{equation}
\begin{equation}\label{TASdotA}
T_{\rm A}\dot{S}_{\rm A}=\frac{\tilde{r}_{\rm A}}{8G}(1-3w_{\rm
eff})\Big[3H(1+w_{\rm eff})F(\phi)+\dot{F}(\phi)\Big],
\end{equation}
where $w_{\rm m}=p_{\rm m}/\rho_{\rm m}$ is the EoS parameter of the
ordinary matter. In the framework of Einstein gravity, i.e. $F{\rm
(\phi)}=E{\rm (\phi)}=1$ and $Z(\phi)=U(\phi)=0$, for a flat FRW
universe dominated by a single fluid ($H^2=\rho_f/3$) filling the
volume enclosed by the apparent (Hubble) horizon ($\tilde{r}_{\rm
A}=H^{-1}$) we have $w_{\rm eff}=w_{\rm m}=w_f$ and Eq.
(\ref{TASdotm}) reduces to Eq. (11) in \cite{Radicella2}.

In the absence of chameleon scalar field, i.e. $E(\phi)=1$, Eq.
(\ref{TASdotm}) reduces to
\begin{equation}\label{TASdotm1}
T_{\rm A}\dot{S}_{\rm m}=\frac{H\tilde{r}^3_{\rm A}\rho_{\rm
m}}{4G}(1+w_{\rm m})(1+3w_{\rm eff}).
\end{equation}
Note that for the ordinary matter we have $w_{\rm m}\geq 0$, hence
the contribution of the matter entropy in the GSL will be positive
or nil for $w_{\rm eff}\geq-1/3$ and negative otherwise.

For a chameleon scalar field minimally coupled to the Ricci scalar
curvature, i.e. $F(\phi)=1$, Eq. (\ref{TASdotA}) yields
\begin{equation}\label{TASdotA1}
T_{\rm A}\dot{S}_{\rm A}=\frac{3H\tilde{r}_{\rm A}}{8G}(1+w_{\rm
eff})(1-3w_{\rm eff}),
\end{equation}
which shows that for $-1\leq w_{\rm eff}\leq 1/3$ the horizon
entropy has a positive or nil contribution in the GSL.

Adding Eqs. (\ref{TASdotm}) and (\ref{TASdotA}) gives the GSL as
\begin{eqnarray}\label{TASdottot}
T_{\rm A}\dot{S}_{\rm tot}=\frac{\tilde{r}_{\rm
A}}{8G}\left\{(1-3w_{\rm eff})\Big[3H(1+w_{\rm
eff})F(\phi)+\dot{F}(\phi)\Big]\right.~~~~~~~\nonumber\\\left.+2\tilde{r}^2_{\rm
A}\rho_{\rm m}(1+w_{\rm m})\left[H(1+3w_{\rm
eff})-\frac{1}{2}\frac{\dot{E}(\phi)}{E(\phi)}\right]\right\}.
\end{eqnarray}
In the Einstein gravity, i.e. $F{\rm (\phi)}=E{\rm (\phi)}=1$ and
$Z(\phi)=U(\phi)=0$, we have $\tilde{r}^2_{\rm A}\rho_{\rm
m}=\tilde{r}^2_{\rm A}\rho_{\rm eff}=3$ and $w_{\rm m}=w_{\rm eff}$
and then the GSL (\ref{TASdottot}) reduces to
\begin{equation}\label{TASdottotE}
T_{\rm A}\dot{S}_{\rm tot}=\frac{9H\tilde{r}_{\rm A}}{8G}(1+w_{\rm
eff})^2\geq 0,
\end{equation}
which is always respected. This also can be obtained by replacing
Eqs. (\ref{ra}) and (\ref{weff1}) into (\ref{GSL-R}).

Note that in general one cannot explore the validity of the GSL
(\ref{TASdottot}) in terms of $w_{\rm eff}$, explicitly, even in
some special cases like $E(\phi)=1$ and $F(\phi)=1$. To do so, we
need to solve the set of Eqs. (\ref{Fr 2}), (\ref{evolu}) and
(\ref{ec}), numerically, for obtaining the evolutions of $a(t)$ (or
$H$), $\phi$ and $\rho_{\rm m}$.

In what follows, we are interested in examining the validity of GSL
for some viable scalar-tensor gravity models. We further assume the
Universe to be spatially flat, i.e. $K=0$, which is compatible with
the recent observations \cite{Komatsu}.

\section{Model I: Brans-Dicke gravity}\label{Model_I}

The action of Brans-Dicke (BD) theory is given by \cite{BD,Arik}\\
\begin{equation}
 I=\int {\rm d}^4x\sqrt{-g}\left[\frac{1}{2k^{2}}\left(\phi R-\frac{\omega}{\phi}
 g^{\mu\nu}\partial_{\mu}\phi\partial_{\nu}\phi\right)+L_{\rm m}\right],\label{action1}
\end{equation}
where $\omega$ is the dimensionless BD parameter.

By comparing the actions (\ref{action1}) and (\ref{action}) one can
get
\begin{equation}
F{(\phi)}=\phi,~~~Z(\phi)=\frac{\omega}{\phi}
,~~~U(\phi)=0,~~~E(\phi)=1.\label{fR}
\end{equation}
Substituting the above relations into Eqs. (\ref{evolu}) and
(\ref{ec}), for a flat universe $(K=0)$ one can obtain
\begin{equation}
\ddot{\phi}+3H\dot{\phi}=\frac{3\phi}{\omega}\left(\dot{H}+2H^{2}\right)+\frac{~\dot{\phi}^{2}}{2\phi},\label{phi
BD1}
\end{equation}
\begin{equation}
\dot{\rho}_{\rm m}+3H(\rho_{\rm m}+p_{\rm m})=0.\label{rom BD1}
\end{equation}
Also the GSL (\ref{S Tot}) for a flat universe $(K=0)$ reduces to
\begin{equation} T_{\rm A}\dot{S}_{\rm
tot}=\frac{2\pi}{H^{4}}\left\{2\phi\dot{H}^{2}-\dot{\phi}\dot{H}H
+2\Big(\dot{H}+ H^{2}
\Big)\Big(\ddot{\phi}+\frac{\omega\dot{\phi}^{2}}{\phi}\Big)\right\},\label{s
tot1}
\end{equation}
where we take $k^2=8\pi G=1$. Equation (\ref{rom BD1}) for the
pressureless matter (i.e. $p_{\rm m}=0$) yields
\begin{equation}
\rho_{\rm m}=\rho_{\rm
m_0}\left(\frac{a}{a_{0}}\right)^{-3}.\label{rom111}
\end{equation}
Replacing Eqs. (\ref{fR}) and (\ref{rom111}) into the second
Friedmann equation (\ref{Fr 2}) yields
\begin{equation}
-2\phi\dot{H}=\rho_{\rm
m_0}\left(\frac{a}{a_{0}}\right)^{-3}+\omega\frac{\dot{\phi}^2}{\phi}
+\ddot{\phi}-H\dot{\phi}.\label{fr2m1}
\end{equation}
To obtain the evolutionary behavior of the GSL (\ref{s tot1}), we
first need to know the time evolution of both the Hubble parameter
$H(t)$ and the scalar field $\phi(t)$. To do so, one can obtain
$a(t)$ and $\phi(t)$ by the numerical solving of Eqs. (\ref{phi
BD1}) and (\ref{fr2m1}). Taking $\Omega_{\rm m_0}=\rho_{\rm
m_0}/(3H_0^2)=0.27$ \cite{2}, $\omega=1.2$ \cite{Farajollahi1} and
using the initial values $a(1)=1$, $\dot{a}(1)=0.84$, $\phi(1)=1.5$
and $\dot{\phi}(1)=1$ \cite{Farajollahi1}, the variations of the
scale factor, the Hubble parameter and the scalar field versus
redshift $z=\frac{a_0}{a}-1$ are plotted in Figs. \ref{fig1a},
\ref{fig1b} and \ref{fig1c}, respectively. Also the evolutionary
behaviors of the deceleration parameter,
\begin{equation}
q=-1-\frac{\dot{H}}{H^2},\label{q}
\end{equation}
and the effective EoS parameter $w_{\rm eff}$, Eq. (\ref{weff1}), in
terms of redshift are plotted in Figs. \ref{fig1d} and \ref{fig1e},
respectively.

Figures \ref{fig1a} to \ref{fig1e} show that: (i) the scale factor,
the Hubble parameter and the scalar field, respectively, increases,
decreases and increases during history of the Universe. (ii) The
deceleration parameter shows a cosmic deceleration $q>0$ to
acceleration $q<0$ transition in the near past which is compatible
with the observations \cite{Ishida}. (iii) The effective EoS
parameter $w_{\rm eff}$ at late times $z\rightarrow -1$ goes to
$-0.6$ which behaves like the quintessence model \cite{Caldwell2}.

With the help of numerical results obtained for the Hubble parameter
and the scalar field presented in Figs. \ref{fig1b} and \ref{fig1c},
the variation of the GSL (\ref{s tot1}) versus $z$ is plotted in
Fig. \ref{fig1f}. The figure shows that the GSL in the BD gravity
model (\ref{action1}) is satisfied during the late cosmological
history of the Universe, i.e. $T_{\rm A}\dot{S}_{\rm tot}\geq 0$.

\section{Model II: BD gravity with a self interacting potential}\label{Model_II}

The action of BD theory with a self interacting potential and a
matter field is given by \cite{Sen}
\begin{equation}\label{action2}
 I=\int {\rm d}^4x\sqrt{-g}\left[\frac{1}{2k^{2}}\left(\phi R-\frac{\omega}{\phi}
 g^{\mu\nu}\partial_{\mu}\phi\partial_{\nu}\phi-V(\phi)\right)+L_{\rm m}\right],
\end{equation}
with
\begin{equation}\label{potential}
V{\rm(\phi)}=\lambda \phi^{4}-\frac{\mu_{0}^{2}}{a(t)^{n}}~\phi^{2}.
\end{equation}
Here $\lambda$ and $\mu_0$ are two constants and $n$ is a positive
integer.

By comparing the actions (\ref{action2}) and (\ref{action}) we find
\begin{equation}
 F(\phi)=\phi,~~~
 Z(\phi)=\frac{\omega}{\rm \phi},~~~U(\phi)=\frac{V(\phi)}{2},~~~E(\phi)=1.\label{f BDV}
\end{equation}
Inserting the above relations into Eqs. (\ref{evolu}) and
(\ref{ec}), for a flat universe one can obtain
\begin{equation}
\frac{2\omega}{\phi}(\ddot{\phi}+3H\dot{\phi})=6(\dot{H}+2H^2)+\omega\left(\frac{\dot{\phi}}{\phi}\right)^2
-4\lambda\phi^3+\mu_0^2\frac{\phi}{a^n}\left(2-n\frac{\dot{a}/a}{\dot{\phi}/\phi}\right),\label{phi
BDV1}
\end{equation}
\begin{equation}
\dot{\rho}_{\rm m}+3H(\rho_{\rm m}+p_{\rm m})=0,\label{rom BDV1}
\end{equation}
where the evolution of $\rho_{\rm m}$ for the pressureless matter
($p_{\rm m}=0$) is same as that obtained in (\ref{rom111}).

Here the GSL (\ref{S Tot}) for a spatially flat universe takes the
form
\begin{equation}\label{s tot2}
T_{\rm A}\dot{S}_{\rm tot}=\frac{2\pi}{
H^{4}}\left\{\dot{H}\Big(2\dot{H}\phi-H\dot{\phi}\Big)
+2\Big(\dot{H}+ H^{2}
\Big)\left(\ddot{\phi}+\frac{\omega}{\phi}\dot{\phi}^{2}\right)\right\},
\end{equation}
where $k^2=8\pi G=1$.

Inserting Eqs. (\ref{rom111}) and (\ref{f BDV}) into the second
Friedmann equation (\ref{Fr 2}), one can find
\begin{equation}
-2\phi\dot{H}=\rho_{\rm
m_0}\left(\frac{a}{a_{0}}\right)^{-3}+\omega\frac{\dot{\phi}^2}{\phi}
+\ddot{\phi}-H\dot{\phi},\label{fr2m2}
\end{equation}
which is same as Eq. ({\ref{fr2m1}) for the BD gravity model
(\ref{action1}). Because the form of potential of the scalar field
does not appear explicitly in the second Friedmann equation (\ref{Fr
2}).

Taking $\Omega_{\rm m_0}=0.27$ \cite{2}$, \omega=1.2$
\cite{Farajollahi1} and $n=1$ \cite{Sen}, the time evolution of both
the scale factor $a(t)$ and the scalar field $\phi(t)$ can be
obtained by numerical solving of Eqs. (\ref{phi BDV1}) and
(\ref{fr2m2}) with the initial values $a(1)=1$, $\dot{a}(1)=0.84$,
$\phi(1)=-1.5$ and $\dot{\phi}(1)=1$. Also we set $\lambda=H_0^2$
and $\mu_0=H_0$ to recast the differential Eqs. (\ref{phi BDV1}) and
(\ref{fr2m2}) in dimensionless form which is more suitable for
numerical integration. The numerical results obtained for $a$, $H$,
$\phi$, $q$ and $w_{\rm eff}$ are plotted in Figs. \ref{fig2a},
\ref{fig2b}, \ref{fig2c}, \ref{fig2d} and \ref{fig2e}, respectively.
Figures show that (i) $a$, $H$ and $\phi$, respectively, increases,
decreases and increases during history of the Universe. (ii) The
deceleration parameter shows a transition from the deceleration era
$q>0$ to the acceleration regime $q<0$. At late times ($z\rightarrow
-1$), the deceleration parameter approaches a de Sitter regime (i.e.
$q\rightarrow-1$), as expected. (iii) The effective EoS parameter
shows a transition from the quintessence state, $w_{\rm eff}>-1$, to
the phantom regime, $w_{\rm eff}<-1$, in the future. Also at late
times we get $w_{\rm eff}\rightarrow-1$ which acts like the
$\Lambda$CDM model.

The results of $H$ and $\phi$ illustrated in Figs. \ref{fig2b} and
\ref{fig2c} help us to obtain the variation of the GSL (\ref{s
tot2}) for the BD gravity model (\ref{action2}) with a self
interacting potential (\ref{potential}). The result is plotted in
Fig. \ref{fig2f}. The figure shows that the GSL for our model is
satisfied from the past to the present epoch. But in the future the
GSL is violated for $z<-0.15$.

\section{Model III: chameleon gravity}\label{Model_III}

The action of chameleon gravity in the presence of matter is given
by \cite{Farajollahi,Farajollahi2}
\begin{equation}
I=\int {\rm
d}^4x\sqrt{-g}\left[\frac{1}{2k^{2}}\Big(R-\label{action3}
 g^{\mu\nu}\partial_{\mu}\phi\partial_{\nu}\phi-2V{(\phi)}\Big)+f{ (\phi)}L_{\rm m}\right],
\end{equation}
where there is a non-minimal coupling term, $f(\phi)L_{\rm m}$,
between the chameleon scalar field and the matter field.

Comparing the chameleon gravity action (\ref{action3}) with action
(\ref{action}), one can get
\begin{equation}\label{fr3}
 F(\phi)=1,~~~Z(\phi)=1,~~~U(\phi)=V(\phi),~~~E(\phi)=f(\phi).
\end{equation}
With the help of these relations, Eqs. (\ref{evolu}) and (\ref{ec})
for a flat universe read
\begin{equation}
\ddot{\phi}+3H\dot{\phi}+V_{,\phi}+\frac{1}{4}(\rho_{\rm m}-3 p_{\rm
m})f_{,\phi}=0,\label{phi 3}
\end{equation}
\begin{equation}
\dot{\rho}_{\rm m}+3H\left( \rho_{\rm m}+p_{\rm m}\right)
=-\frac{3}{4}\frac{\dot{f}(\phi)}{f(\phi)}(\rho_{\rm m}+p_{\rm
m}),\label{ec5}
\end{equation}
where we take $k^2=8\pi G=1$. Taking the integration of Eq.
(\ref{ec5}) for the pressureless matter ($p_{\rm m}=0$) gives
\begin{equation}
\rho_{\rm m}=\rho_{\rm m_0}\left(\frac{a}{a_{0}}\right)^{-3}
\left(\frac{f(\phi)}{f_{0}}\right)^{-\frac{3}{4}}.\label{omega3}
\end{equation}
Substituting the relations (\ref{fr3}) into (\ref{S Tot}) gives the
GSL for a flat universe as
\begin{equation}\label{s tot3}
T_{\rm A}\dot{S}_{\rm
tot}=\frac{2\pi}{H^{4}}\left\{-2\dot{H}\Big(2H^{2}
+\dot{H}\Big)+\frac{1}{2f(\phi)}\Big(2\dot{H}+
\dot{\phi}^{2}\Big)\left(4\Big
(H^{2}+\dot{H}\Big)+\frac{\dot{f}}{f}H\right)\right\}.
\end{equation}
According to \cite{Farajollahi2} we consider both $f(\phi)$ and
potential $V(\phi)$ appeared in (\ref{action3}) behave exponentially
as
\begin{equation}\label{mod3}
f(\phi)=f_{0}e^{b_{1}\phi},~~~V(\phi)=V_{0}e^{b_{2}\phi},
\end{equation}
where $f_0$, $V_0$, $b_1$ and $b_2$ are arbitrary constants.

Using Eqs. (\ref{omega3}) and (\ref{mod3}), the evolution Eq.
(\ref{phi 3}) for the pressureless matter ($p_{\rm m}=0$) gives
\begin{equation}\label{phi 33}
\ddot{\phi}+3H\dot{\phi}+b_2V_0e^{b_2\phi}+\frac{1}{4}b_1f_0\rho_{\rm
m_0}\left(\frac{a}{a_{0}}\right)^{-3}e^{b_1\phi/4}=0.
\end{equation}
Also the second Friedmann equation (\ref{Fr 2}) reads
\begin{equation}\label{fr2m3}
-2\dot{H}=f_0\rho_{\rm
m_0}\left(\frac{a}{a_{0}}\right)^{-3}e^{b_1\phi/4}+\dot{\phi}^2.
\end{equation}
Finally, the GSL (\ref{s tot3}) takes the form
\begin{equation}\label{s totcc 55}
T_{\rm A}\dot{S}_{\rm
tot}=\frac{2\pi}{H^{4}}\left\{-2\dot{H}\Big(2H^{2} +\dot{H}\Big)
+\frac{1}{2f_{0}e^{b_{1}\phi}}\Big(2\dot{H}+\dot{\phi}^2\Big)
\Big(4\Big(H^{2}+\dot{H}\Big)+b_{1}\dot{\phi}H\Big)\right\}.
\end{equation}
From numerical solving of Eqs. (\ref{phi 33}) and (\ref{fr2m3}) one
can obtain $a(t)$ and $\phi(t)$. To do so we take $\Omega_{\rm
m_0}=0.27$ \cite{2}, $f_0=-10$ and $b_1=b_2=-1$ \cite{Farajollahi2}
and use the initial values $a(1)=1$, $\dot{a}(1)=1$, $\phi(1)=1$ and
$\dot{\phi}(1)=-2$. The variations of $a$, $H$, $\phi$, $q$ and
$w_{\rm eff}$ versus redshift are plotted in Figs. \ref{fig3a} to
\ref{fig3e}. Figures show that (i) the scale factor increases when
the time increases. The Hubble parameter decreases with increasing
time and then increases to approach a constant value. The scalar
field decreases with increasing time and increases at late times.
(ii) The deceleration parameter shows a cosmic transition from $q>0$
to $q<0$ in the near past which is compatible with the observations
\cite{Ishida}. It also approaches a de Sitter regime at late times,
as expected. (iii) The effective EoS parameter can justify the
transition from the quintessence state ($w_{\rm eff}>-1$) to the
phantom regime ($w_{\rm eff}<-1$) in the near past, as indicated by
recent observations \cite{Sahni}. This is also in good agreement
with that obtained in \cite{Farajollahi2}. The effective EoS
parameter also behaves like the $\Lambda$CDM model at late times.

With the help of numerical results obtained for the Hubble parameter
and the scalar field illustrated in Figs. \ref{fig3b} and
\ref{fig3c}, the variation of the GSL (\ref{s totcc 55}) versus
redshift for the chameleon gravity model is plotted in Fig.
\ref{fig3f}. The figure shows that the GSL in this model is violated
for the range of $-0.88<z<0.37$. This is in contrast with that
obtained in \cite{Farajollahi2}. Authors of Ref. \cite{Farajollahi2}
investigated the GSL in flat FRW chameleon cosmology and showed that
in an expanding universe, the GSL is always respected. This
contradiction comes back to the definition of energy and pressure in
the Gibbs equation (\ref{gi}). In \cite{Farajollahi2}, Farajollahi
et al. considered effective (total) energy $E_{\rm eff}=\rho_{\rm
eff}V$ and pressure $p_{\rm eff}$ instead of $E_{\rm m}=\rho_{\rm
m}V$ and $p_{\rm m}$ as we have in our case. Therefore in
\cite{Farajollahi2}, the GSL is defined as $T_{\rm A}\dot{S}_{\rm
tot}=T_{\rm A}(\dot{S}_{\rm eff}+\dot{S}_{\rm A})$ in which $$T_{\rm
A}\dot{S}_{\rm eff}=\frac{3H\tilde{r}_{\rm A}}{4G}(1+w_{\rm
eff})(1+3w_{\rm eff}),$$ and $T_{\rm A}\dot{S}_{\rm A}$ is given by
Eq. (\ref{TASdotA1}). Finally the GSL yields Eq. (\ref{TASdottotE})
which is nothing but the GSL in the Einstein gravity. This confirms
that the GSL investigated in \cite{Farajollahi2} does not belong to
the chameleon gravity.
\section{Model IV: chameleonic generalized BD gravity}\label{Model_IV}

The action of chameleonic generalized BD gravity model is given by
\cite{Farajollahi1,Pavon}
\begin{equation}\label{action4}
I=\int d^4x\sqrt{-g}\left[\frac{1}{2k^{2}}\left(\phi
R-\frac{\omega(\phi)}{\phi}
 g^{\mu\nu}\partial_{\mu}\phi\partial_{\nu}\phi\right)+f(\phi)L_{\rm m}\right].
\end{equation}

Comparing Eq. (\ref{action4}) with action (\ref{action}) gives
\begin{equation}\label{CGBD4}
 F(\phi)=\phi,~~~Z(\phi)=\frac{\omega(\phi)}{\rm
 \phi},~~~U(\phi)=0,~~~E(\phi)=f(\phi).
\end{equation}
Using these, Eqs. (\ref{evolu}) and (\ref{ec}) for a flat universe
read
\begin{equation}\label{phi BDc V4}
\ddot{\phi}+3H\dot{\phi}=\frac{1}{2\omega(\phi)+3}\left[\Big(\rho_{\rm
m}-3p_{\rm m}\Big)\left(f(\phi)-\frac{1}{2}\phi
f_{,\phi}\right)-\omega_{,\phi}\dot{\phi}^{2} \right],
\end{equation}
\begin{equation}
\dot{\rho}_{\rm m}+3H\left( \rho_{\rm m}+p_{\rm m}\right)
=-\frac{3}{4}\frac{\dot{f}(\phi)}{f(\phi)}(\rho_{\rm m}+p_{\rm
m}),\label{ec3}
\end{equation}
where $k^2=8\pi G=1$. Solution of Eq. (\ref{ec3}) for the
pressureless matter ($p_{\rm m}=0$) yields the same result obtained
in (\ref{omega3}).

With the help of relations (\ref{CGBD4}), the GSL (\ref{S Tot}) for
a flat universe yields
\begin{eqnarray}\label{stot5}
T_{\rm A}\dot{S}_{\rm tot}=\frac{2\pi}{H^{4}}\left\{\Big(2H^{2}
+\dot{H}\Big)\Big( \dot{\phi} H -2\phi\dot{H}\Big)+\frac{1}{2f
(\phi)}\left(4\Big(H^{2}+\dot{H}\Big)+H\frac{\dot{f}}{f}
\right)\right.\nonumber\\\left.\times\left(2\phi\dot{H}-\dot{\phi}H+\ddot{\phi}
+\omega(\phi)\frac{\dot{\phi}^{2}}{\phi} \right)\right\}.
\end{eqnarray}
According to \cite{Farajollahi1} we take
\begin{equation}\label{mod4}
f(\phi)=f_{0}e^{b\phi},~~~\omega(\phi)=\omega_{ 0}{\phi}^{n}.
\end{equation}

Substituting Eqs. (\ref{omega3}) and (\ref{mod4}) into (\ref{phi BDc
V4}) for the pressureless matter ($p_{\rm m}=0$) gives
\begin{equation}\label{phi BDcg V44}
\ddot{\phi}+3H\dot{\phi}=\frac{1}{3+2\omega_0\phi^n}\left[f_0\rho_{\rm
m_0}\left(\frac{a}{a_{0}}\right)^{-3}\left(1-\frac{b}{2}\phi\right)e^{b\phi/4}-n\omega_0\phi^{n-1}\dot{\phi}^2\right].
\end{equation}
Also the second Friedmann equation (\ref{Fr 2}) gives
\begin{equation}\label{fr2m4}
-2\phi\dot{H}=f_0\rho_{\rm
m_0}\left(\frac{a}{a_{0}}\right)^{-3}e^{b\phi/4}+\omega_0\phi^{n-1}\dot{\phi}^2+\ddot{\phi}-H\dot{\phi}.
\end{equation}
Moreover the GSL (\ref{stot5}) reduces to
\begin{eqnarray}\label{s tot33}
T_{\rm A}\dot{S}_{\rm tot}=\frac{2\pi}{H^{4}}\left\{\Big(2H^{2}
+\dot{H}\Big)\Big( \dot{\phi} H
-2\phi\dot{H}\Big)+\frac{1}{2f_{0}e^{b\phi}}\left(4\Big(H^{2}+\dot{H}\Big)+b\dot{\phi}H
\right)\right.\nonumber\\\left.\times\left(2\phi\dot{H}-\dot{\phi}H+\ddot{\phi}
+\omega_0\phi^{n-1}\dot{\phi}^{2}\right)\right\}.
\end{eqnarray}
Taking $\Omega_{\rm m_0}=0.27$ \cite{2}$, \omega_0=1.2$, $n=-2$,
$f_0=-7$ and $b=-0.4$ \cite{Farajollahi1}, the time evolution of
both the scale factor $a(t)$ and the scalar field $\phi(t)$ can be
obtained by numerical solving of Eqs. (\ref{phi BDcg V44}) and
(\ref{fr2m4}) with the initial values $a(1)=1$, $\dot{a}(1)=1$,
$\phi(1)=-6.5$ and $\dot{\phi}(1)=0.1$. Figures \ref{fig4a} to
\ref{fig4e} show that: (i) the scale factor and the Hubble
parameter, respectively, increases and decreases with increasing
time. (ii) The scalar field with increasing time, decreases to a
minimum and then increases to approach a constant value. (iii) The
deceleration parameter shows the cosmic transition $q>0\rightarrow
q<0$ in the near past, as indicated by recent observations
\cite{Ishida}. It also approaches a de Sitter regime at late times,
as expected. (iv) The effective EoS parameter at late times behaves
like the $\Lambda$CDM model ($w_{\rm eff}\rightarrow -1$).

The results of $H$ and $\phi$ illustrated in Figs. \ref{fig4b} and
\ref{fig4c} help us to obtain the variation of the GSL (\ref{s
tot33}) for the chameleonic generalized BD gravity model
(\ref{action4}). The result is plotted in Fig. \ref{fig4f}. The
figure shows that the GSL for this model is violated during the late
cosmological history of the Universe.
\section{Model V: chameleonic BD gravity with a self interacting potential}\label{Model_V}

Within the framework of chameleonic BD gravity with a self
interacting potential, the action is given by \cite{Sheykhi3}
\begin{equation}\label{action5}
I=\int {\rm d}^4x\sqrt{-g}\left[\frac{1}{2k^{2}}\left(\phi
R-\frac{\omega}{\phi}
 g^{\mu\nu}\partial_{\mu}\phi\partial_{\nu}\phi-V(\phi)\right)+f (\phi)L_{\rm m}\right].
\end{equation}

Here in comparison with action (\ref{action}) we have
\begin{equation}\label{CBD}
 F(\phi)=\phi,~~~Z(\phi)=\frac{\omega}{ \phi},
~~~U(\phi)=\frac{V(\phi)}{2},~~~E(\phi)=f(\phi).
\end{equation}
Using the above relations, the evolution Eq. (\ref{evolu}) and the
continuity Eq. (\ref{ec}) for a flat universe take the forms
\begin{equation}\label{phi BDc V1}
\ddot{\phi}+3H\dot{\phi}=\frac{1}{2\omega+3}\left[(\rho_{\rm
m}-3p_{\rm m})\left(f(\phi)-\frac{1}{2}\phi
f_{,\phi}\right)+2V(\phi)-\phi V_{,\phi}\right],
\end{equation}
\begin{equation}
\dot{\rho}_{\rm m}+3H\left( \rho_{\rm m}+p_{\rm m}\right)
=-\frac{3}{4}\frac{\dot{f}(\phi)}{f(\phi)}(\rho_{\rm m}+p_{\rm
m}),\label{ec4}
\end{equation}
where $k^2=8\pi G=1$. Solution of Eq. (\ref{ec4}) for the
pressureless matter ($p_{\rm m}=0$) is same as that obtained in
(\ref{omega3}).

Replacing the relations (\ref{CBD}) into (\ref{S Tot}) gives the GSL
for the chameleonic BD gravity with a self interacting potential as
\begin{eqnarray}\label{s tot5}
T_{\rm A}\dot{S}_{\rm tot}=\frac{2\pi}{H^{4}}\left\{\Big(2H^{2}
+\dot{H}\Big)\Big( \dot{\phi} H -2\phi\dot{H}\Big)+\frac{1}{2f
(\phi)}\left(4\Big(H^{2}+\dot{H}\Big)+H\frac{\dot{f}}{f}
\right)\right.\nonumber\\\left.\times\left(2\phi\dot{H}-\dot{\phi}H+\ddot{\phi}
+\omega\frac{\dot{\phi}^{2}}{\phi} \right)\right\}.
\end{eqnarray}
According to \cite{Khoury} we consider the inverse power-law
potential
\begin{equation}\label{mod5V}
V(\phi)=\frac{M^{n+4}}{\phi^{n}},
\end{equation}
where $M$ has units of mass and $n$ is a positive constant.

We further take \cite{Farajollahi1}
\begin{equation}\label{mod5}
f{\rm (\phi)}=f_{0}e^{b\phi},
\end{equation}
where $f_0$ and $b$ are constant parameters.

Inserting Eqs. (\ref{omega3}), (\ref{mod5V}) and (\ref{mod5}) into
(\ref{phi BDc V1}), for the pressureless matter ($p_{\rm m}=0$) one
can obtain
\begin{eqnarray}\label{phi BDc V2}
\ddot{\phi}+3H\dot{\phi}=\frac{1}{3+2\omega}\left[f_0\rho_{\rm
m_0}\left(\frac{a}{a_{0}}\right)^{-3}\left(1-\frac{b}{2}\phi\right)e^{b\phi/4}+\frac{(n+2)M^{n+4}}{\phi^n}\right].
\end{eqnarray}
Also the second Friedmann equation (\ref{Fr 2}) reduces to
\begin{equation}\label{fr2m5}
-2\phi\dot{H}=f_0\rho_{\rm
m_0}\left(\frac{a}{a_{0}}\right)^{-3}e^{b\phi/4}+\omega\frac{\dot{\phi}^2}{\phi}+\ddot{\phi}-H\dot{\phi}.
\end{equation}
Furthermore the GSL (\ref{s tot5}) yields
\begin{eqnarray}\label{s tot44}
T_{\rm A}\dot{S}_{\rm tot}=\frac{2\pi}{H^{4}}\left\{\Big(2H^{2}
+\dot{H}\Big)\Big( \dot{\phi} H
-2\phi\dot{H}\Big)+\frac{1}{2f_{0}e^{b\phi}}\left(4\Big(H^{2}+\dot{H}\Big)+b\dot{\phi}H
\right)\right.\nonumber\\\left.\times\left(2\phi\dot{H}-\dot{\phi}H+\ddot{\phi}
+\omega\frac{\dot{\phi}^{2}}{\phi} \right)\right\}.
\end{eqnarray}
From Eqs. (\ref{phi BDc V2}) and (\ref{fr2m5}), the scale factor
$a(t)$ and the scalar field $\phi(t)$ can be obtained, numerically.
To do so, we take $\Omega_{\rm m_0}=0.27$ \cite{2}$, \omega=1.2$,
$f_0=-7$, $b=-0.4$ \cite{Farajollahi1}, $n=2$ \cite{Khoury} and use
the initial values $a(1)=1$, $\dot{a}(1)=1$, $\phi(1)=1$ and
$\dot{\phi}(1)=-1.4$. The results are plotted in Fig. \ref{ModelV}.
Figure show that (i) the scale factor is an increasing function of
time , as expected for an expanding universe. (ii) The Hubble
parameter and the scalar field decrease with increasing time,
approach to a minimum in the future and then increase when the time
increases. (iii) The deceleration parameter at late times goes to
$-1$ which acts like the de Sitter model. It also shows a cosmic
transition from $q>0$ to $q<0$ in the future. (iv) The effective EoS
parameter at late times behaves like the $\Lambda$CDM model. It also
shows the phantom divide line crossing in the future.

Using the numerical results obtained for $H$ and $\phi$, the
evolutionary behavior of the GSL (\ref{s tot44}) for the chameleonic
BD gravity with a self interacting potential is plotted in Fig.
\ref{fig5f}. The figure shows that the GSL for this model is
satisfied from the past to the present epoch. But in the future the
GSL is violated for $z<-0.53$.
\section{Conclusions}\label{Conclusions}

Here, we investigated the GSL in the framework of scalar-tensor
gravity. In a general theory of scalar-tensor gravity, a scalar
field can be nonminimally coupled both to the scalar curvature (as
Brans-Dicke field) and the matter Lagrangian (as chameleon field) in
the action. Hence, we extended the action of ordinary scalar-tensor
gravity theory to the case in which there is a non-minimal coupling
between the scalar field and the matter field. Then we derived the
associated filed equations governing the gravity and the scalar
field. For a FRW universe filled with the ordinary matter, we
obtained the modified Friedmann equations as well as the evolution
equation of the scalar field.  We further assumed the boundary of
FRW universe to be enclosed by the dynamical apparent horizon which
is in thermal equilibrium with the Hawking temperature. Then we
obtained a general expression for the GSL of gravitational
thermodynamics. For some viable scalar-tensor gravity models
containing BD gravity, BD gravity with a self interacting potential,
chameleon gravity, chameleonic generalized BD gravity, and
chameleonic BD gravity with a self interacting potential, we first
obtained the evolutionary behaviors of the matter density, the scale
factor, the Hubble parameter, the scalar field, the deceleration
parameter as well as the effective EoS parameter. Then, we examined
the validity of the GSL for the aforementioned models. Our results
show the following.

(i) The aforementioned models can give rise to a late time
accelerated expansion phase for the Universe. The deceleration
parameter for the all models shows a cosmic deceleration $q>0$ to
acceleration $q<0$ transition. In the BD gravity model, the
chameleon gravity model and the chameleonic generalized BD gravity
model, the cosmic transition from $q>0$ to $q<0$ occurs in the near
past which is compatible with the observations \cite{Ishida}. For
all models but the BD gravity model, at late times ($z\rightarrow
-1$), the deceleration parameter approaches a de Sitter regime (i.e.
$q\rightarrow-1$), as expected.

(ii) The effective EoS parameter for the BD gravity model with a
self interacting potential, the chameleon gravity model and the
chameleonic BD gravity model with a self interacting potential,
shows a transition from the quintessence state, $w_{\rm eff}>-1$, to
the phantom regime, $w_{\rm eff}<-1$. For the chameleon gravity
model, the transition from $w_{\rm eff}>-1$ to $w_{\rm eff}<-1$
occurs in the near past, as indicated by recent observations
\cite{Sahni}. For all models but the BD gravity model, the effective
EoS parameter at late times behaves like the $\Lambda$CDM model
($w_{\rm eff}\rightarrow -1$).

(iii) The GSL for the BD gravity model like the Einstein gravity is
satisfied during the late cosmological history of the Universe. For
the BD gravity model with a self interacting potential, the GSL is
satisfied from the past to the present epoch. But in the future the
GSL is violated for $z<-0.15$. For the BD gravity model with/without
a self interacting potential, the contribution of the matter entropy
in the GSL will be positive or nil for $w_{\rm eff}\geq-1/3$ and
negative otherwise. For the chameleon gravity model, the GSL is
violated for the range of $-0.88<z<0.37$. However, for $-1\leq
w_{\rm eff}\leq 1/3$ the horizon entropy has a positive or nil
contribution in the GSL. For the chameleonic generalized BD gravity
model, the GSL is violated during the late cosmological history of
the Universe. Finally, for the chameleonic BD gravity model with a
self interacting potential, the GSL is satisfied from the past to
the present time. But in the future the GSL is violated for
$z<-0.53$.

\subsection*{Acknowledgements}

The authors thank the referee for his/her valuable comments. The
works of A. Abdolmaleki and K. Karami have been supported
financially by Research Institute for Astronomy and Astrophysics of
Maragha (RIAAM) under research project No. 1/3569.


\clearpage
\begin{figure}[h]
\begin{minipage}[b]{1\textwidth}
\subfigure[\label{fig1a} ]{ \includegraphics[width=.39\textwidth]%
{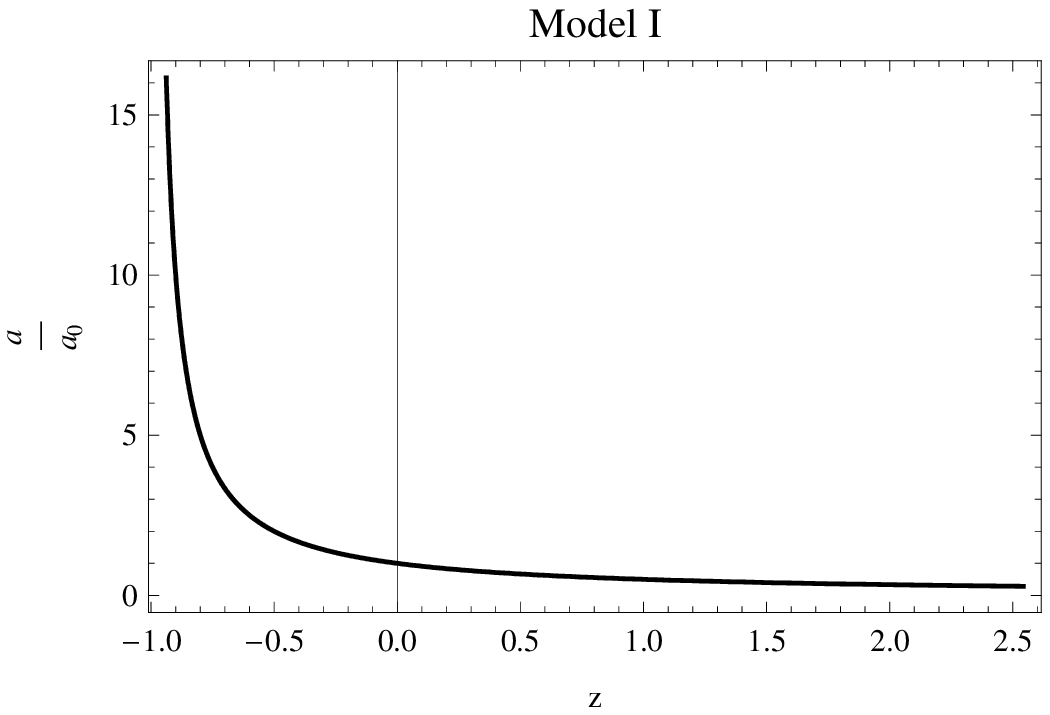}} \hspace{.2cm}
\subfigure[\label{fig1b} ]{ \includegraphics[width=.39\textwidth]%
{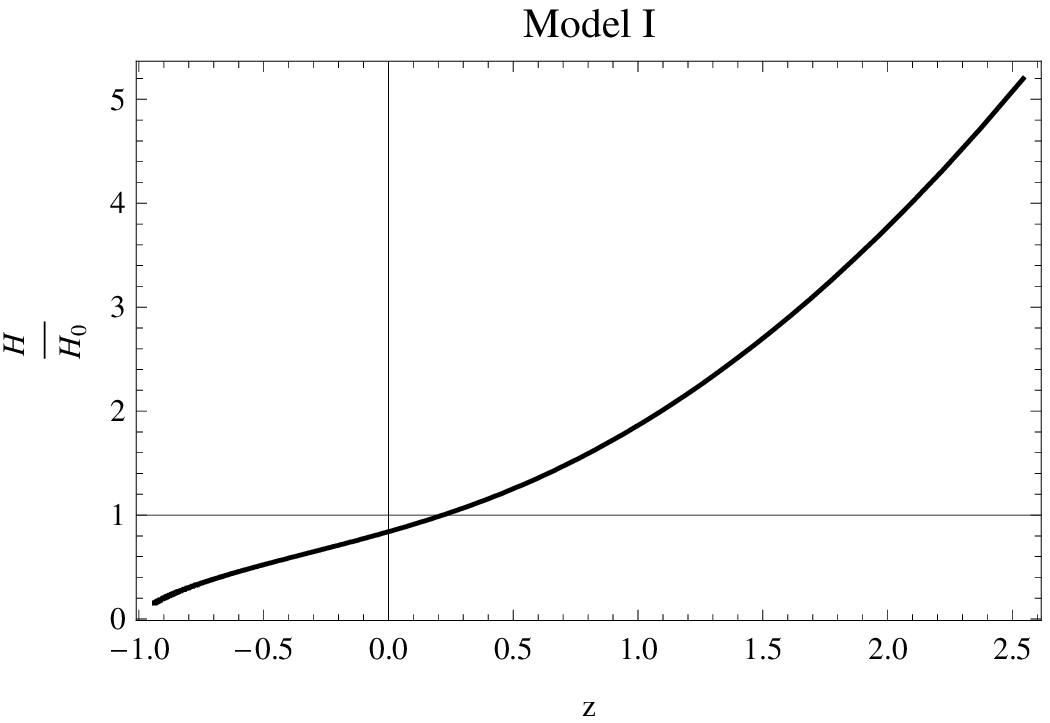}}
\end{minipage}
\\
\begin{minipage}[b]{1\textwidth}
\subfigure[\label{fig1c} ]{ \includegraphics[width=.39\textwidth]%
{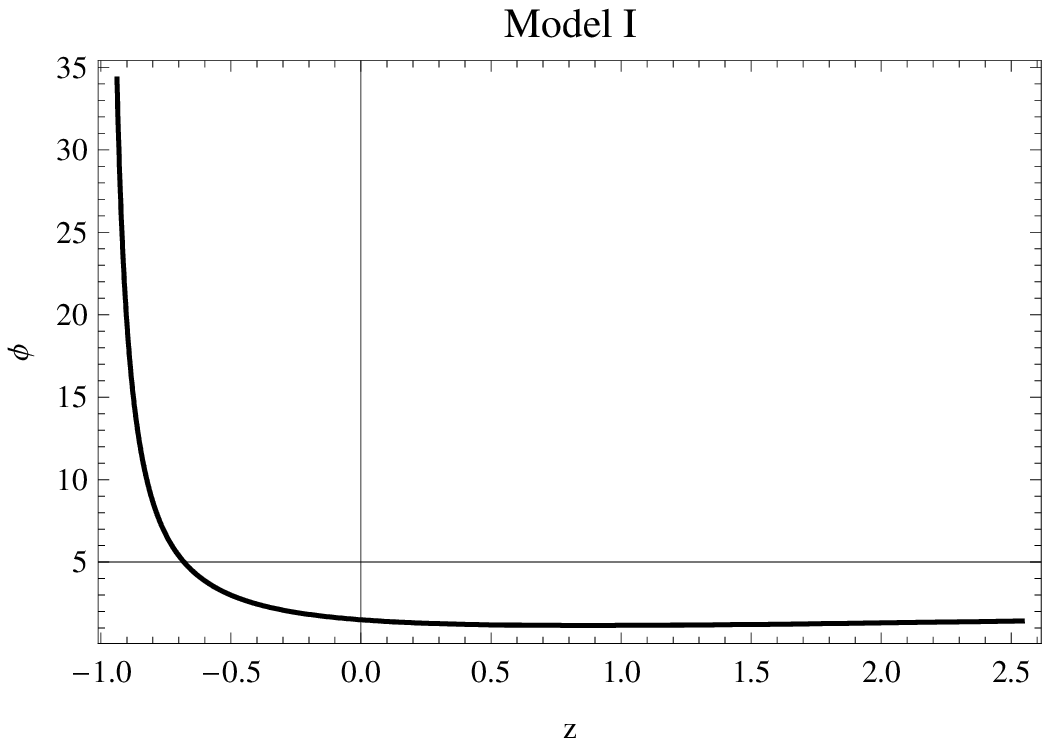}} \hspace{.2cm}
\subfigure[\label{fig1d} ]{ \includegraphics[width=.39\textwidth]%
{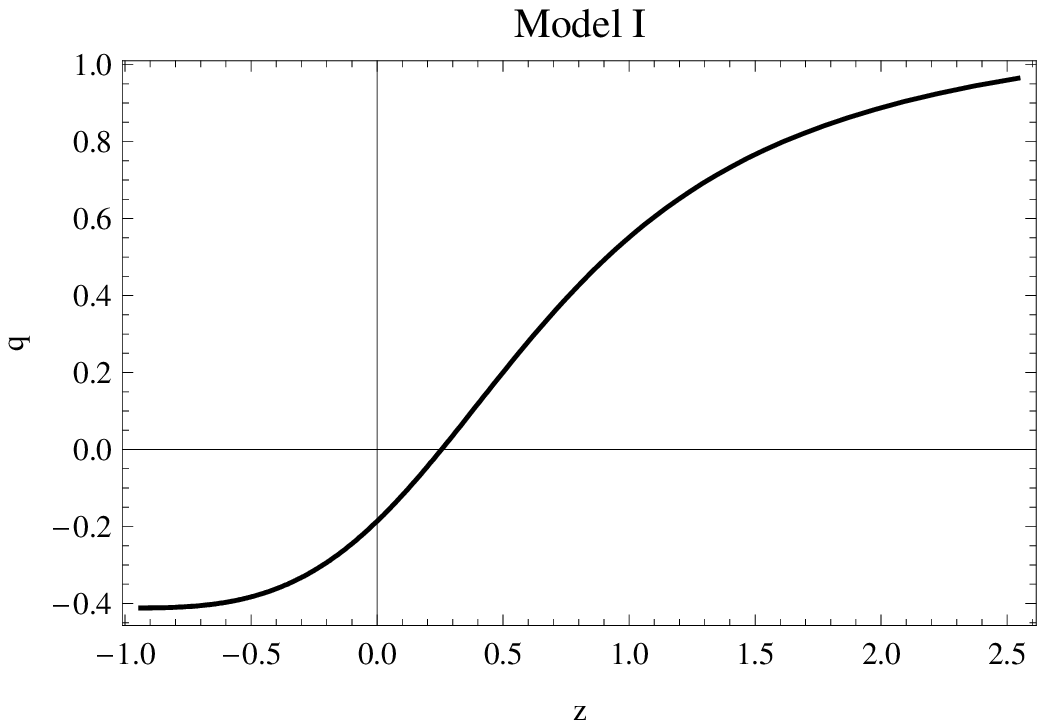}}
\end{minipage}
\\
\begin{minipage}[b]{1\textwidth}
\subfigure[\label{fig1e} ]{ \includegraphics[width=.39\textwidth]%
{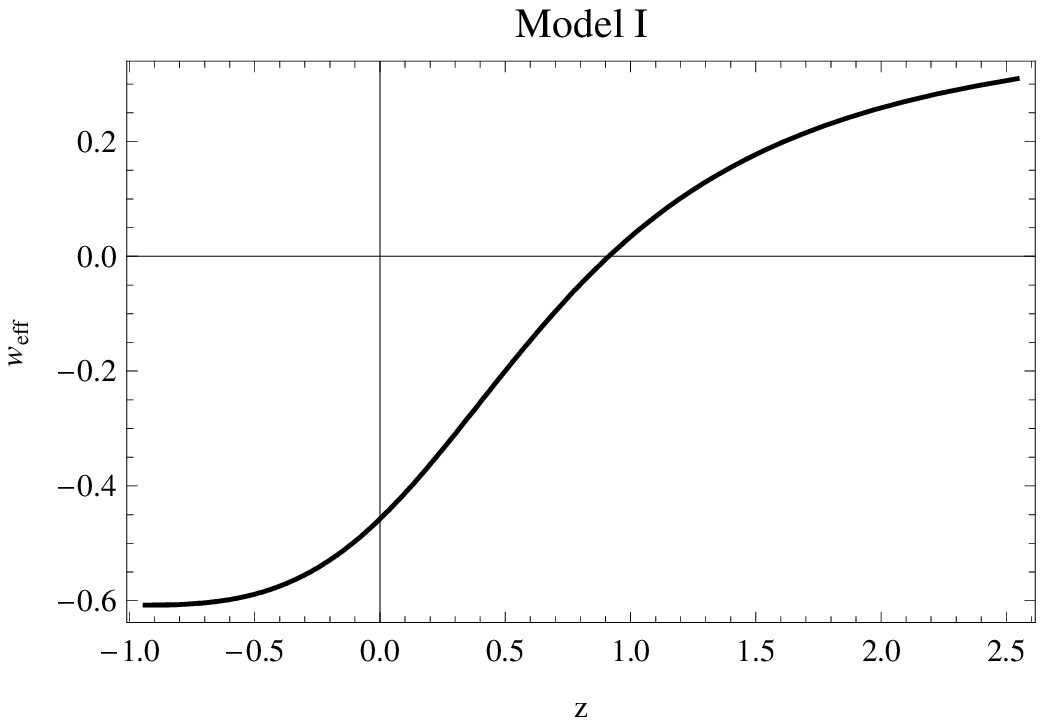}} \hspace{.8cm}
\subfigure[\label{fig1f} ]{ \includegraphics[width=.39\textwidth]%
{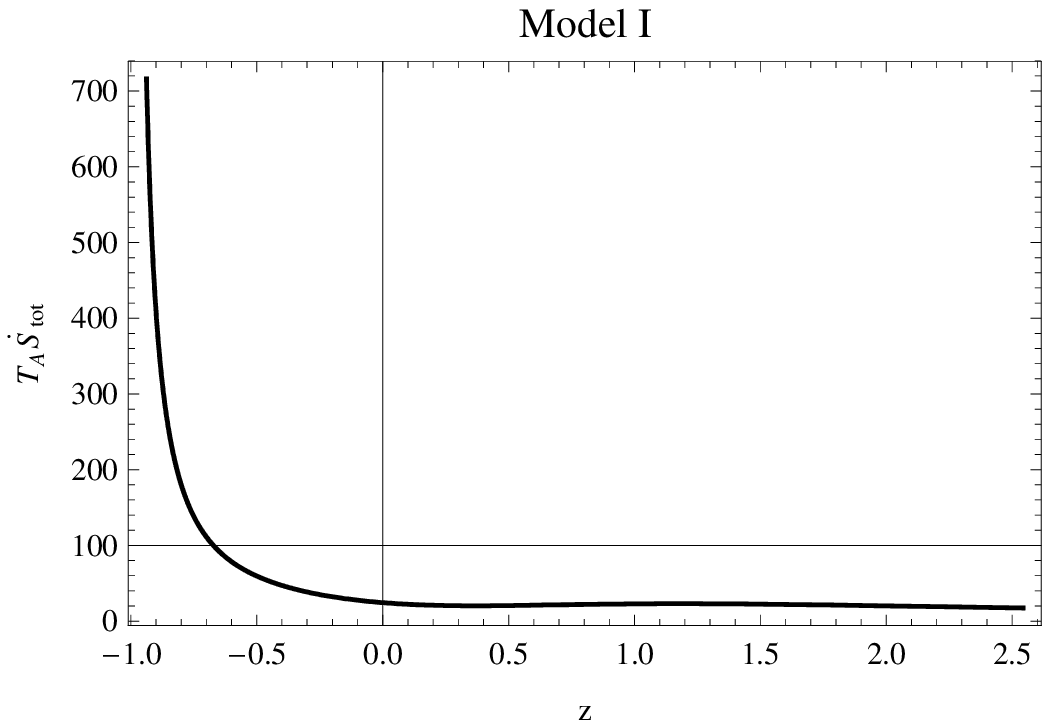}}
\end{minipage}
\caption{The variations of the scale factor $a$, the Hubble
parameter $H$, the scalar field $\phi$, the deceleration parameter
$q$, the effective EoS parameter $w_{\rm eff}$ and the GSL, $T_{\rm
A}\dot{S}_{\rm tot}$, versus redshift $z$ for model I
(\ref{action1}). Initial values are $a(1)=1$, $\dot{a}(1)=0.84$,
$\phi(1)=1.5$ and $\dot{\phi}(1)=1$ \cite{Farajollahi1}. Auxiliary
parameters are: $\Omega_{\rm m_0}=0.27$ \cite{2} and $\omega=1.2$
\cite{Farajollahi1}. Here $t_0=1/H_0$.} \label{ModelI}
\end{figure}
\clearpage
\begin{figure}[h]
\begin{minipage}[b]{1\textwidth}
\subfigure[\label{fig2a} ]{ \includegraphics[width=.39\textwidth]%
{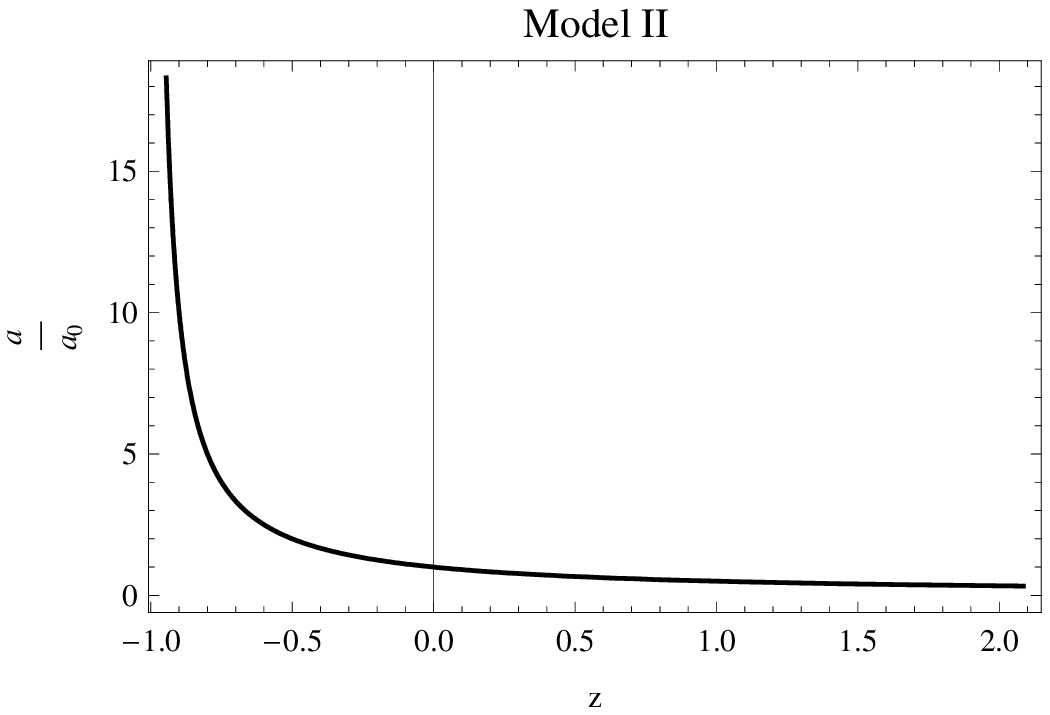}} \hspace{.2cm}
\subfigure[\label{fig2b} ]{ \includegraphics[width=.39\textwidth]%
{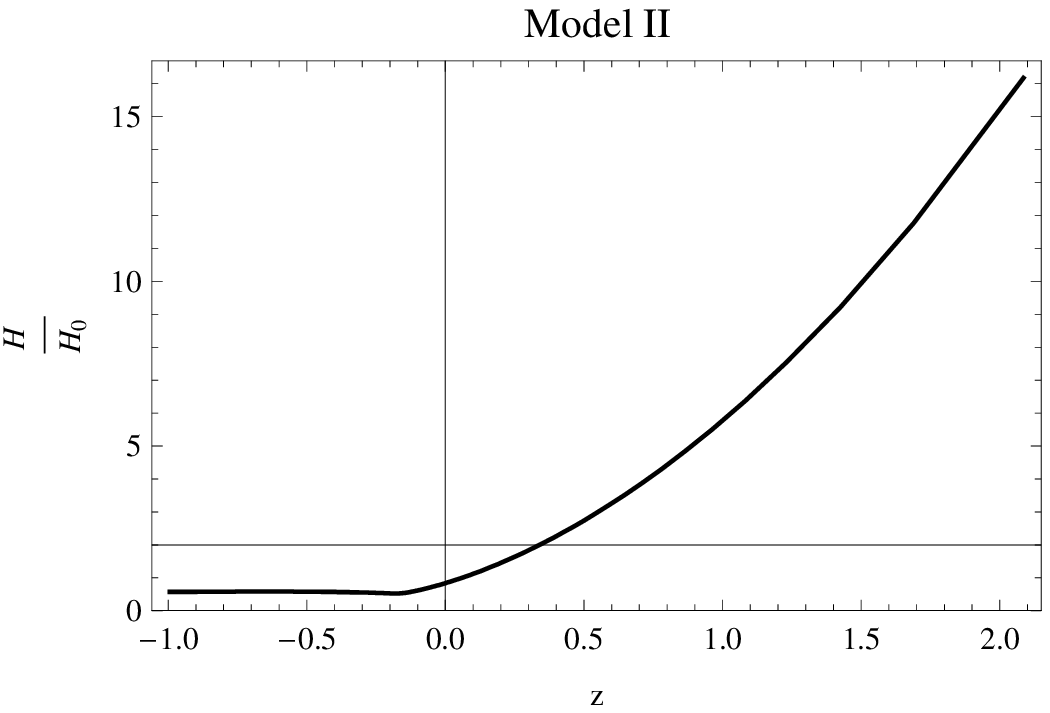}}
\end{minipage}
\\
\begin{minipage}[b]{1\textwidth}
\subfigure[\label{fig2c} ]{ \includegraphics[width=.39\textwidth]%
{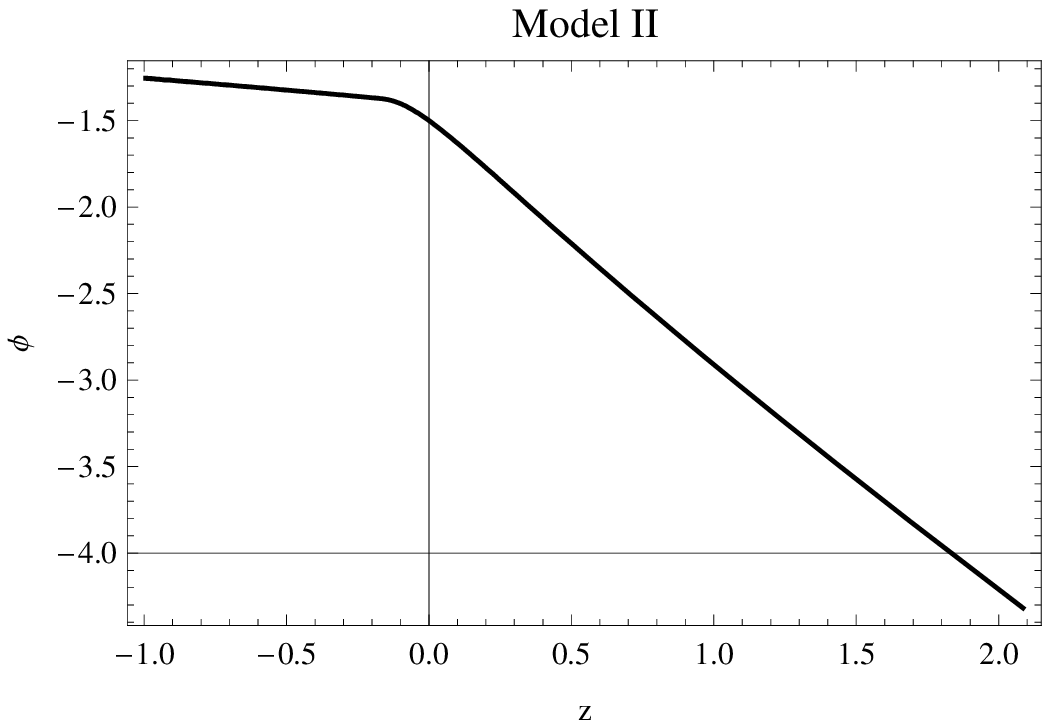}} \hspace{.2cm}
\subfigure[\label{fig2d} ]{ \includegraphics[width=.39\textwidth]%
{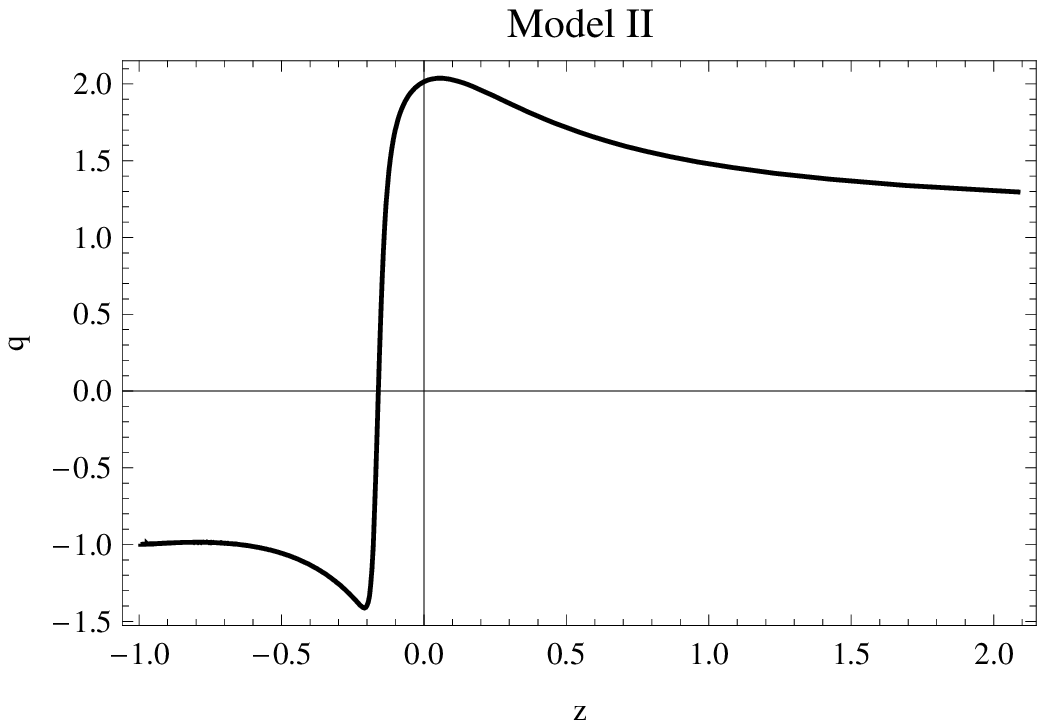}}
\end{minipage}
\\
\begin{minipage}[b]{1\textwidth}
\subfigure[\label{fig2e} ]{ \includegraphics[width=.39\textwidth]%
{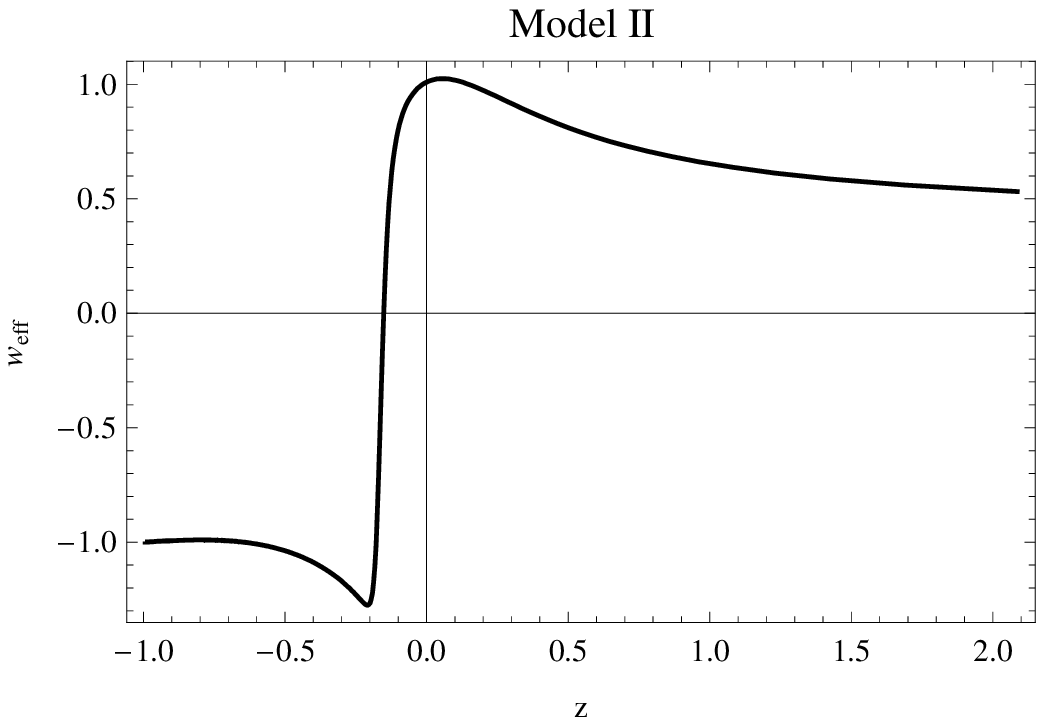}} \hspace{.8cm}
\subfigure[\label{fig2f} ]{ \includegraphics[width=.39\textwidth]%
{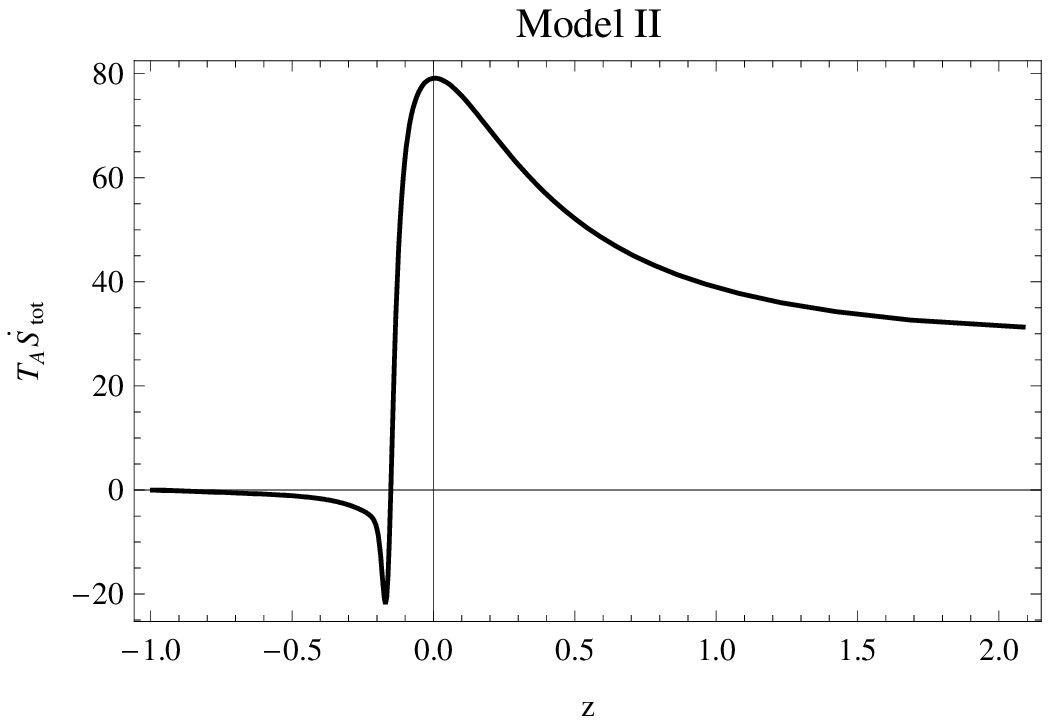}}
\end{minipage}
\caption{Same as Fig. \ref{ModelI} for model II (\ref{action2}).
Initial values are $a(1)=1$, $\dot{a}(1)=0.84$, $\phi(1)=-1.5$ and
$\dot{\phi}(1)=1$. Auxiliary parameters are: $\Omega_{\rm m_0}=0.27$
\cite{2}$, \omega=1.2$ \cite{Farajollahi1} and $n=1$ \cite{Sen}.
Here $t_0=1/H_0$, $\lambda=H_0^2$ and $\mu_0=H_0$.} \label{ModelII}
\end{figure}
\clearpage
\begin{figure}[h]
\begin{minipage}[b]{1\textwidth}
\subfigure[\label{fig3a} ]{ \includegraphics[width=.39\textwidth]%
{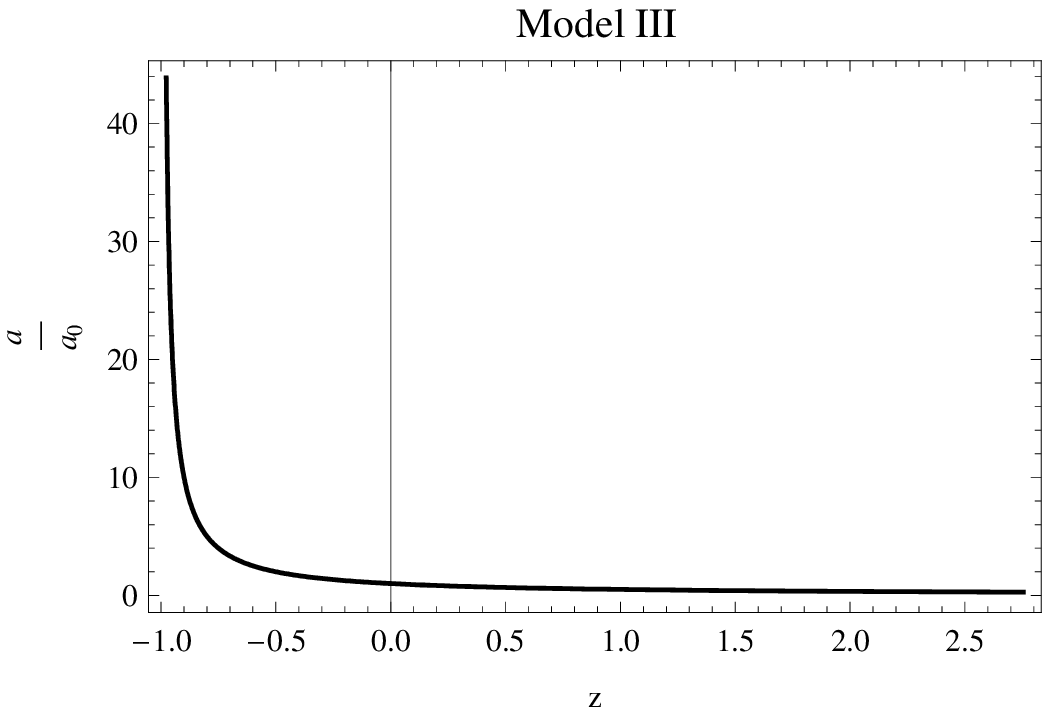}} \hspace{.2cm}
\subfigure[\label{fig3b} ]{ \includegraphics[width=.39\textwidth]%
{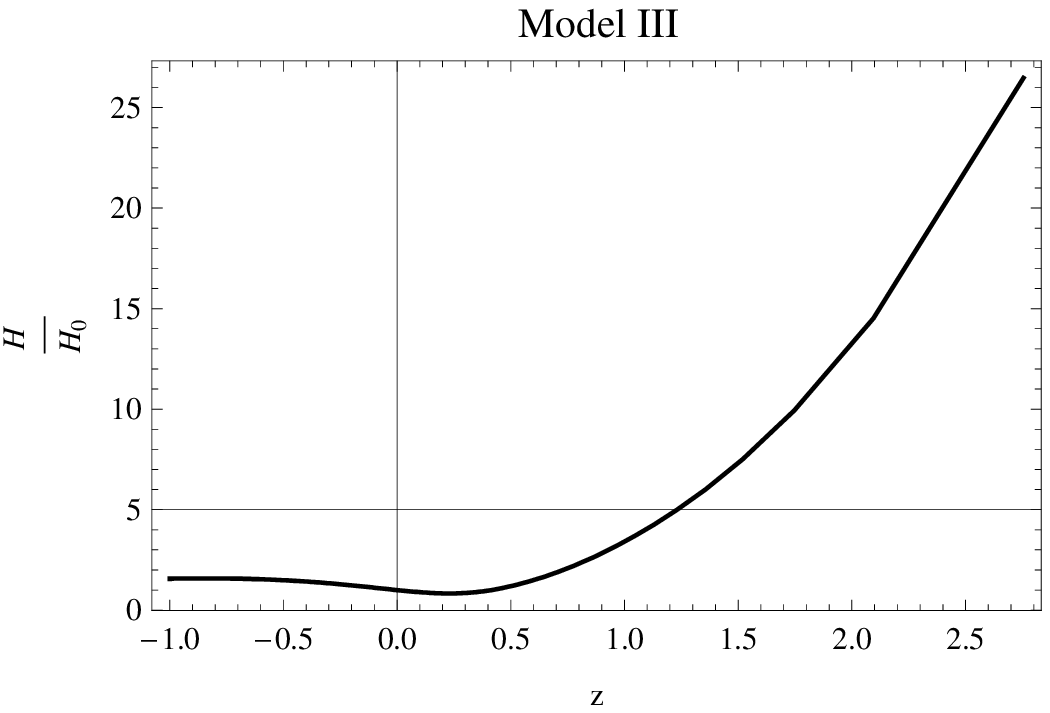}}
\end{minipage}
\\
\begin{minipage}[b]{1\textwidth}
\subfigure[\label{fig3c} ]{ \includegraphics[width=.39\textwidth]%
{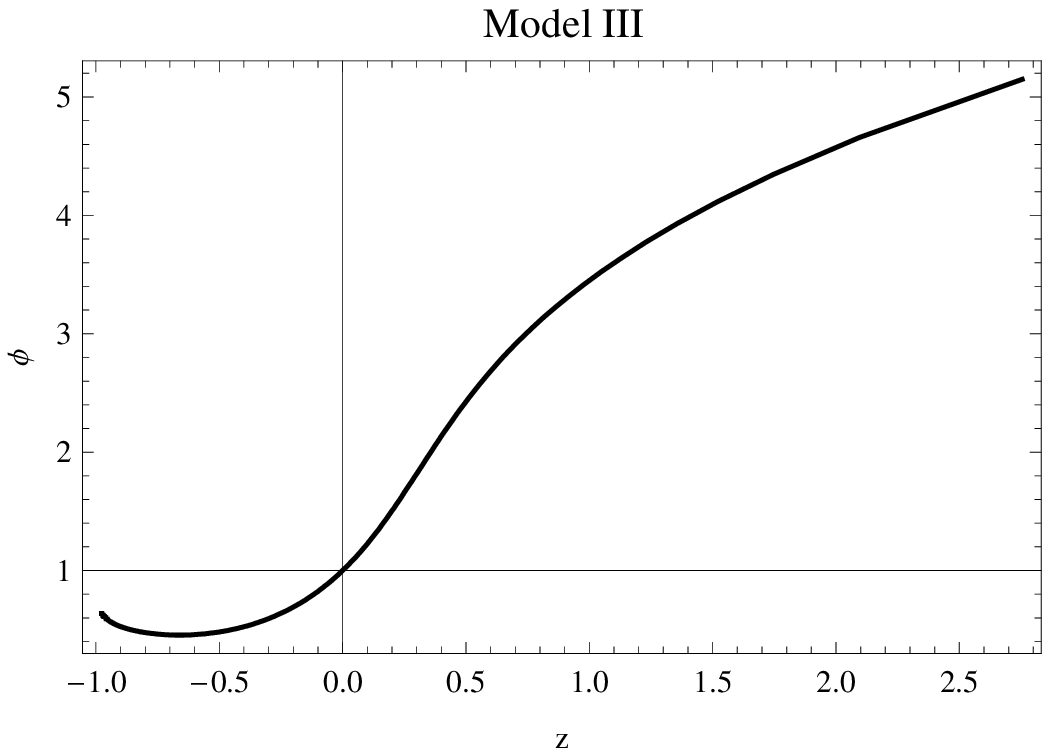}} \hspace{.2cm}
\subfigure[\label{fig3d} ]{ \includegraphics[width=.39\textwidth]%
{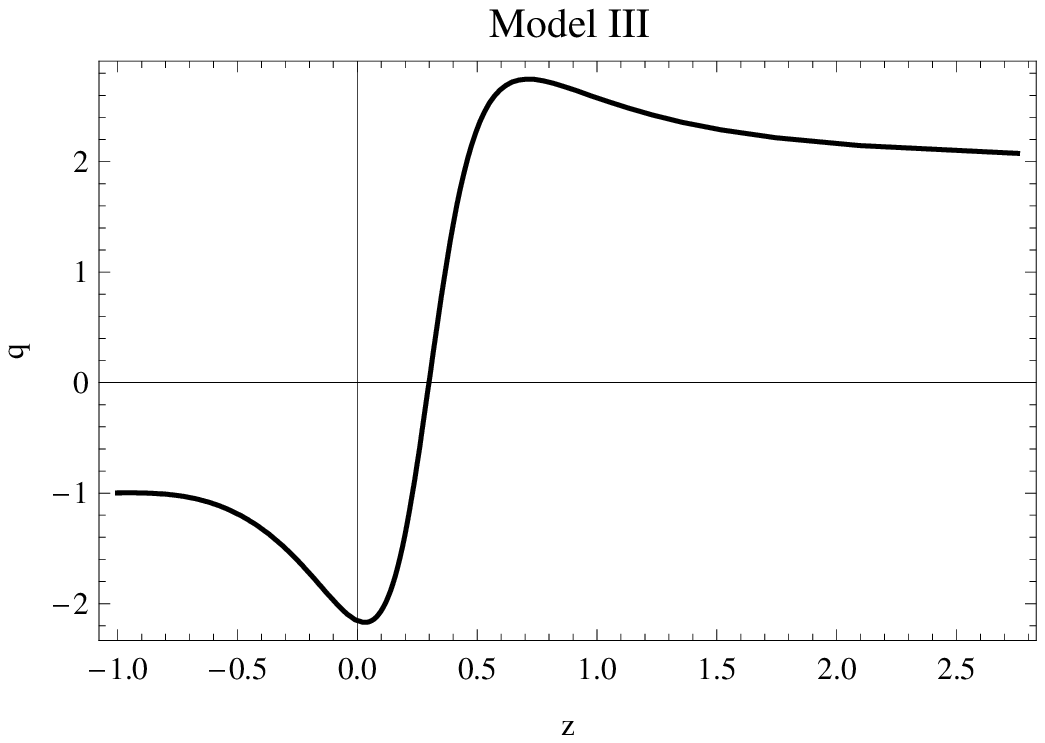}}
\end{minipage}
\\
\begin{minipage}[b]{1\textwidth}
\subfigure[\label{fig3e} ]{ \includegraphics[width=.39\textwidth]%
{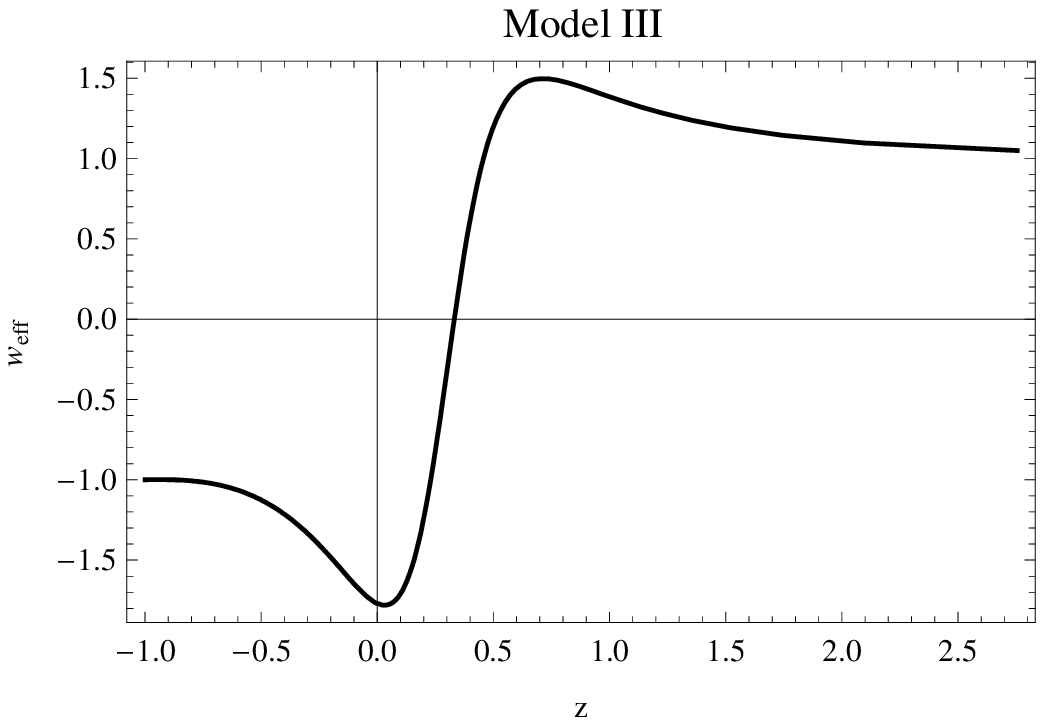}} \hspace{.8cm}
\subfigure[\label{fig3f} ]{ \includegraphics[width=.39\textwidth]%
{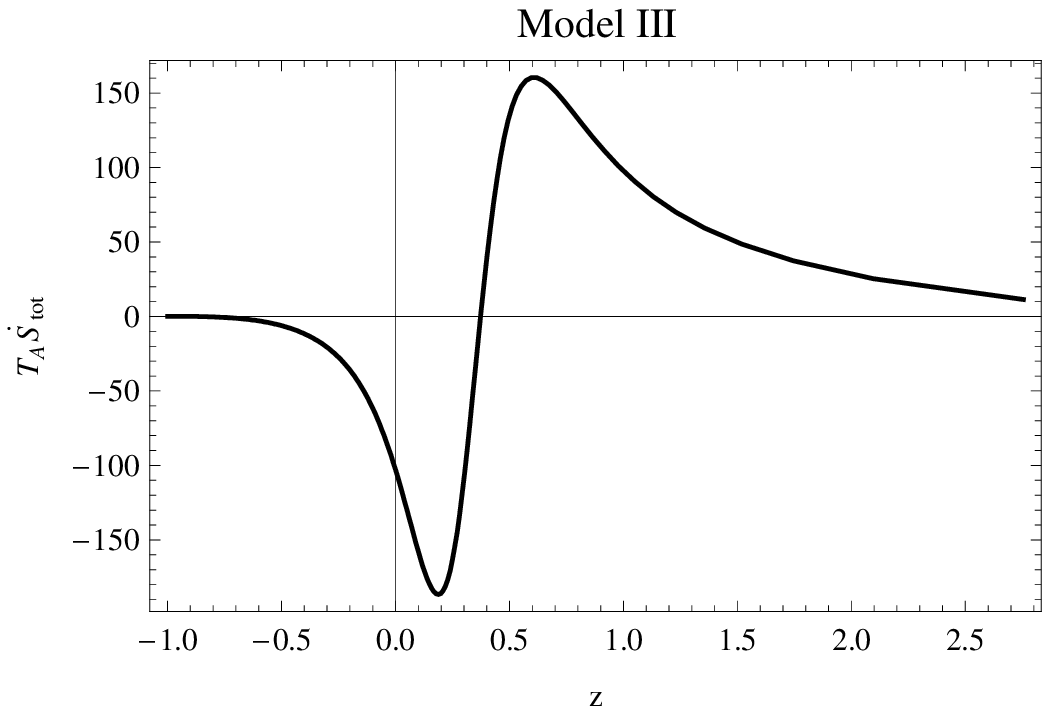}}
\end{minipage}
\caption{Same as Fig. \ref{ModelI} for model III (\ref{action3}).
Initial values are $a(1)=1$, $\dot{a}(1)=1$, $\phi(1)=1$ and
$\dot{\phi}(1)=-2$. Auxiliary parameters are: $\Omega_{\rm
m_0}=0.27$ \cite{2}, $f_0=-10$ and $b_1=b_2=-1$ \cite{Farajollahi2}.
Here $t_0=1/H_0$ and $V_0=H_0^2$.} \label{ModelIII}
\end{figure}
\clearpage
\begin{figure}[h]
\begin{minipage}[b]{1\textwidth}
\subfigure[\label{fig4a} ]{ \includegraphics[width=.39\textwidth]%
{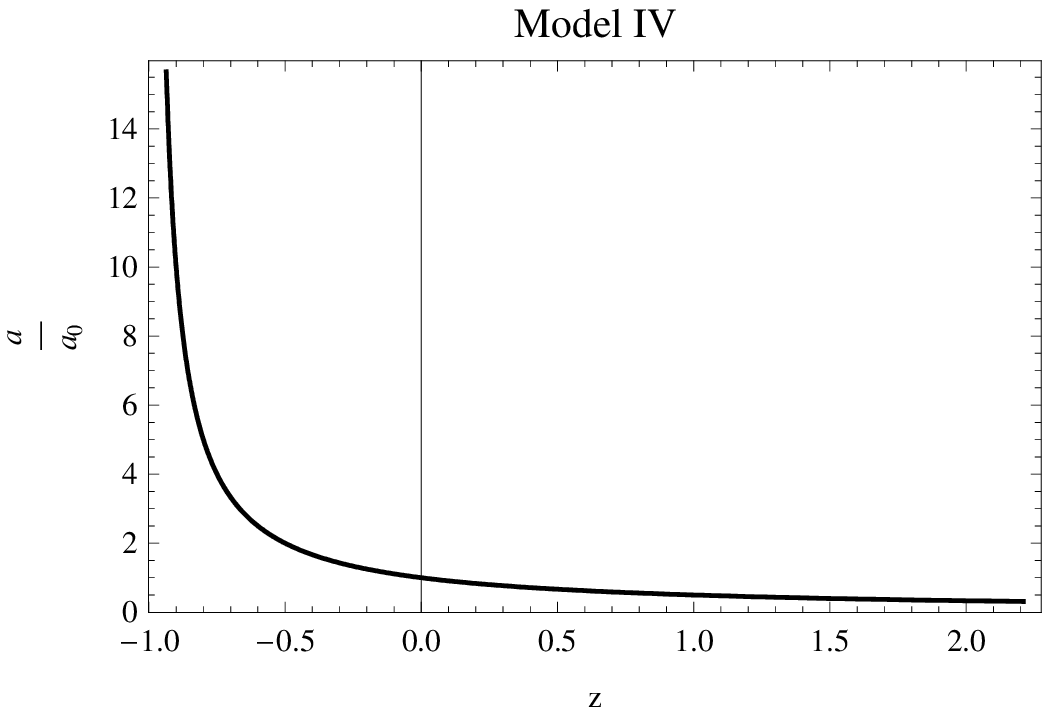}} \hspace{.2cm}
\subfigure[\label{fig4b} ]{ \includegraphics[width=.39\textwidth]%
{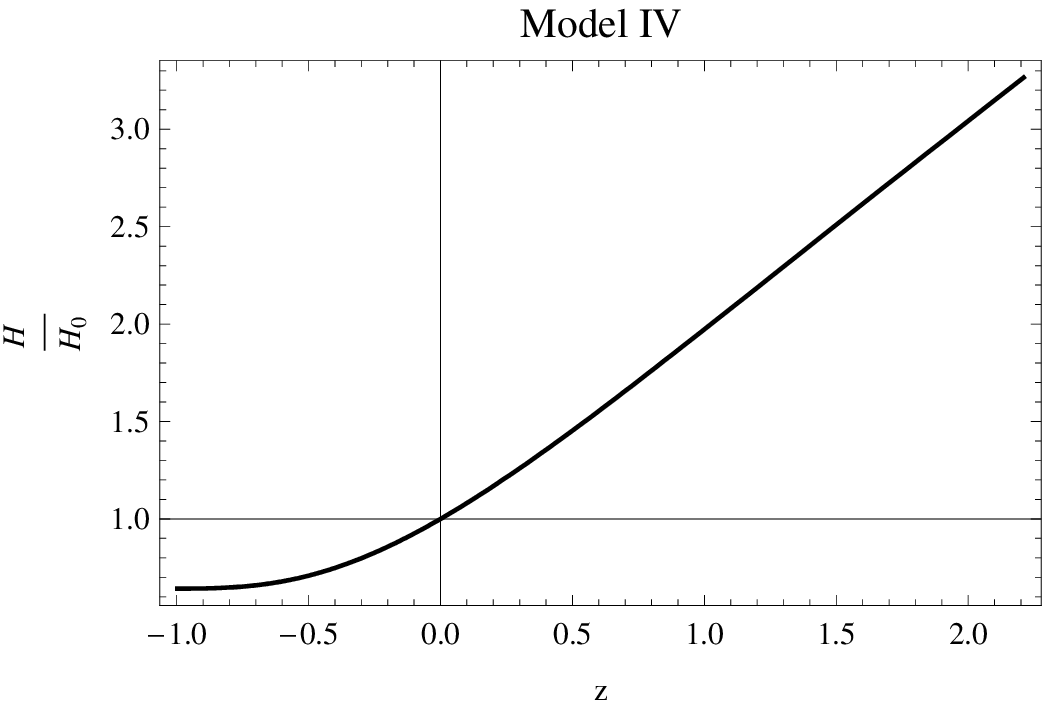}}
\end{minipage}
\\
\begin{minipage}[b]{1\textwidth}
\subfigure[\label{fig4c} ]{ \includegraphics[width=.39\textwidth]%
{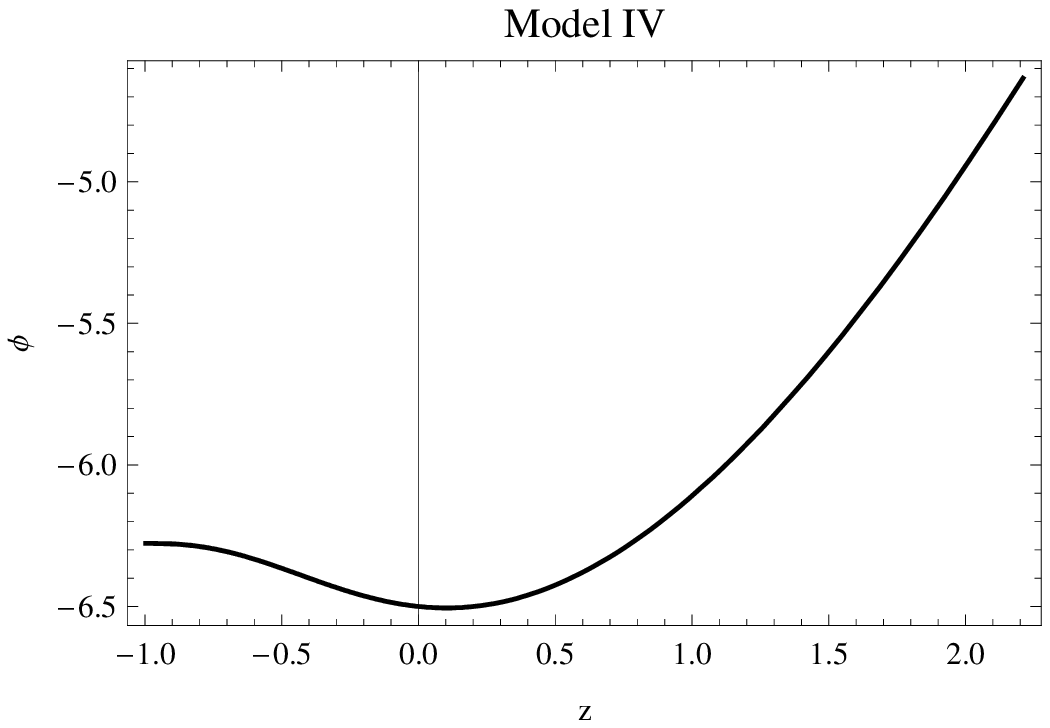}} \hspace{.2cm}
\subfigure[\label{fig4d} ]{ \includegraphics[width=.39\textwidth]%
{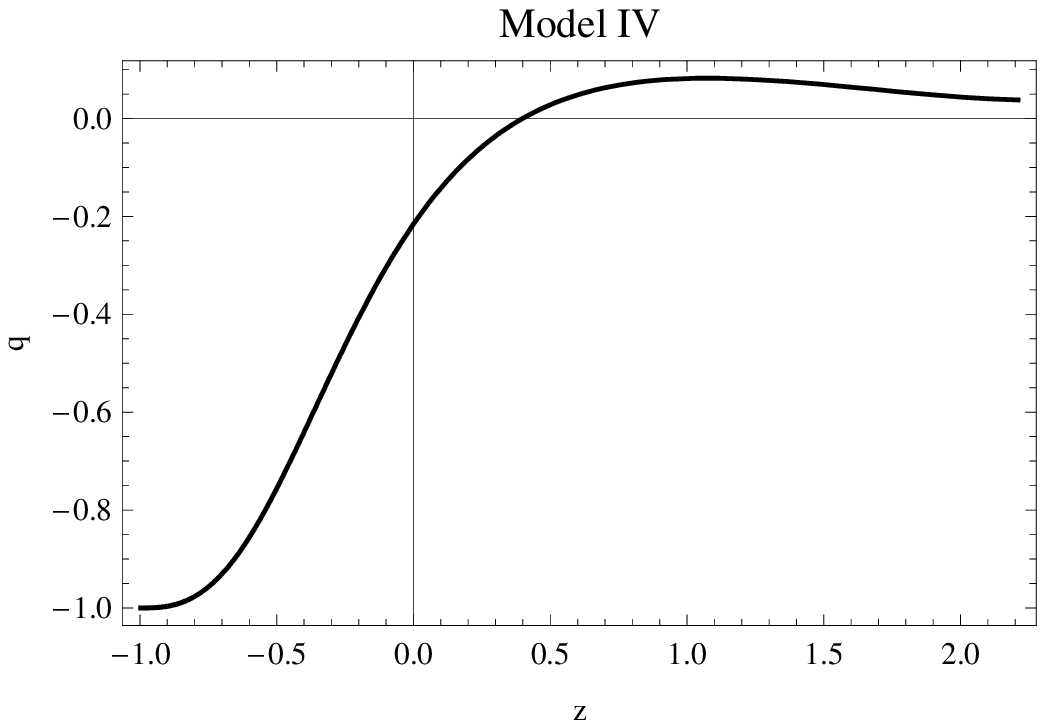}}
\end{minipage}
\\
\begin{minipage}[b]{1\textwidth}
\subfigure[\label{fig4e} ]{ \includegraphics[width=.39\textwidth]%
{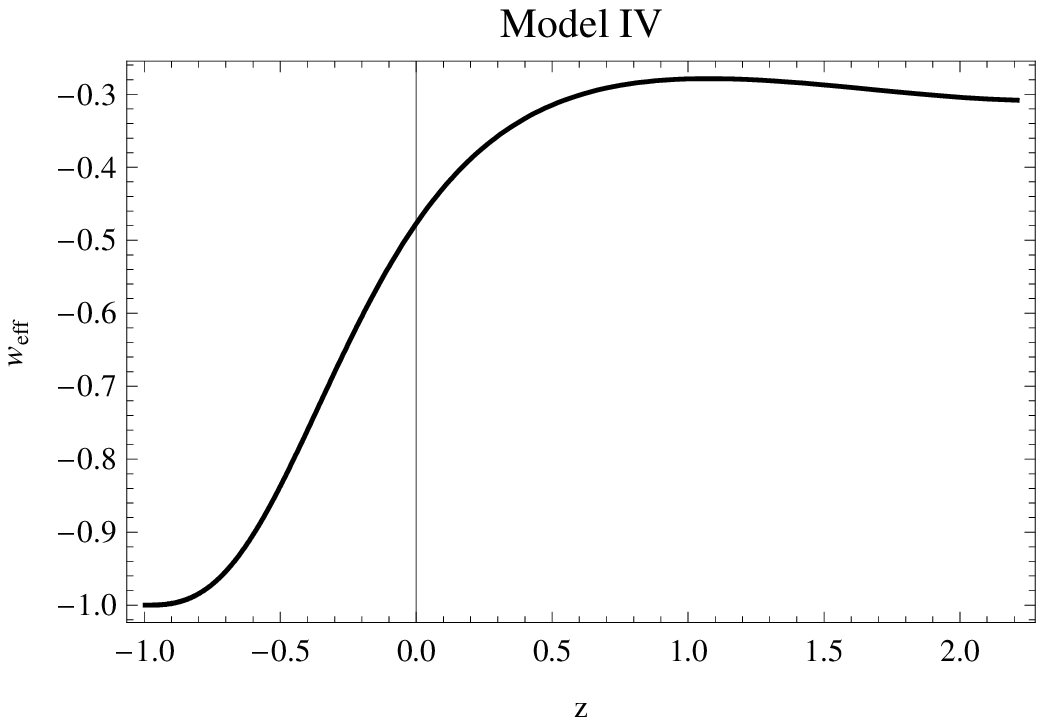}} \hspace{.8cm}
\subfigure[\label{fig4f} ]{ \includegraphics[width=.39\textwidth]%
{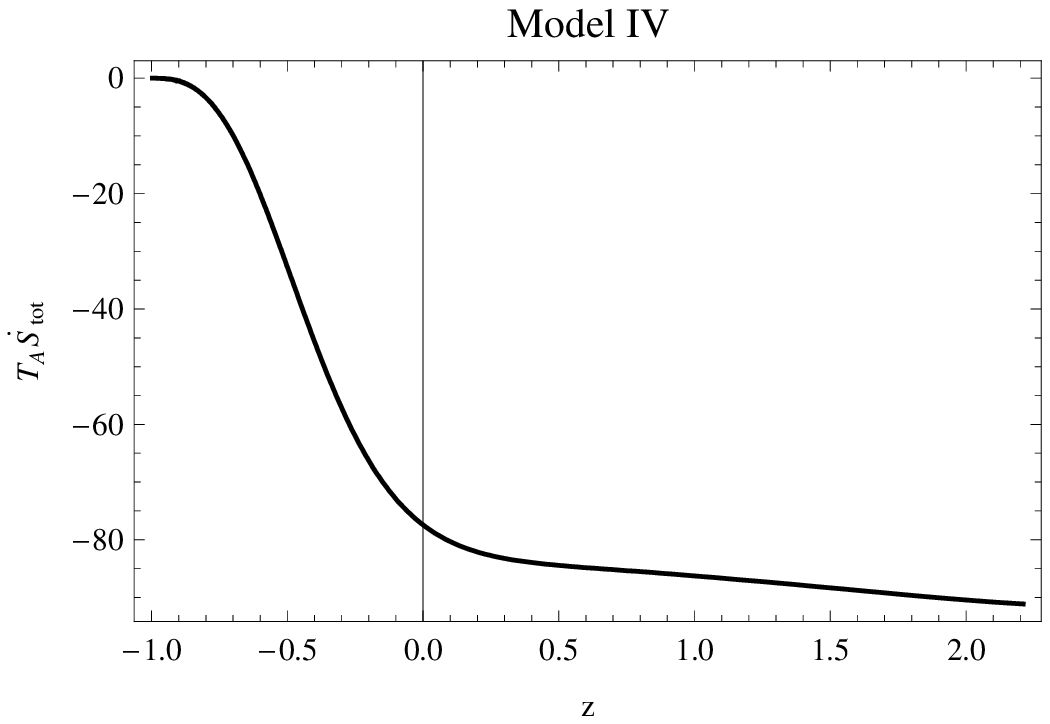}}
\end{minipage}
\caption{Same as Fig. \ref{ModelI} for model IV (\ref{action4}).
Initial values are $a(1)=1$, $\dot{a}(1)=1$, $\phi(1)=-6.5$ and
$\dot{\phi}(1)=0.1$. Auxiliary parameters are: $\Omega_{\rm
m_0}=0.27$ \cite{2}$, \omega_0=1.2$, $n=-2$, $f_0=-7$ and $b=-0.4$
\cite{Farajollahi1} . Here $t_0=1/H_0$.} \label{ModelIV}
\end{figure}
\clearpage
\begin{figure}[h]
\begin{minipage}[b]{1\textwidth}
\subfigure[\label{fig5a} ]{ \includegraphics[width=.39\textwidth]%
{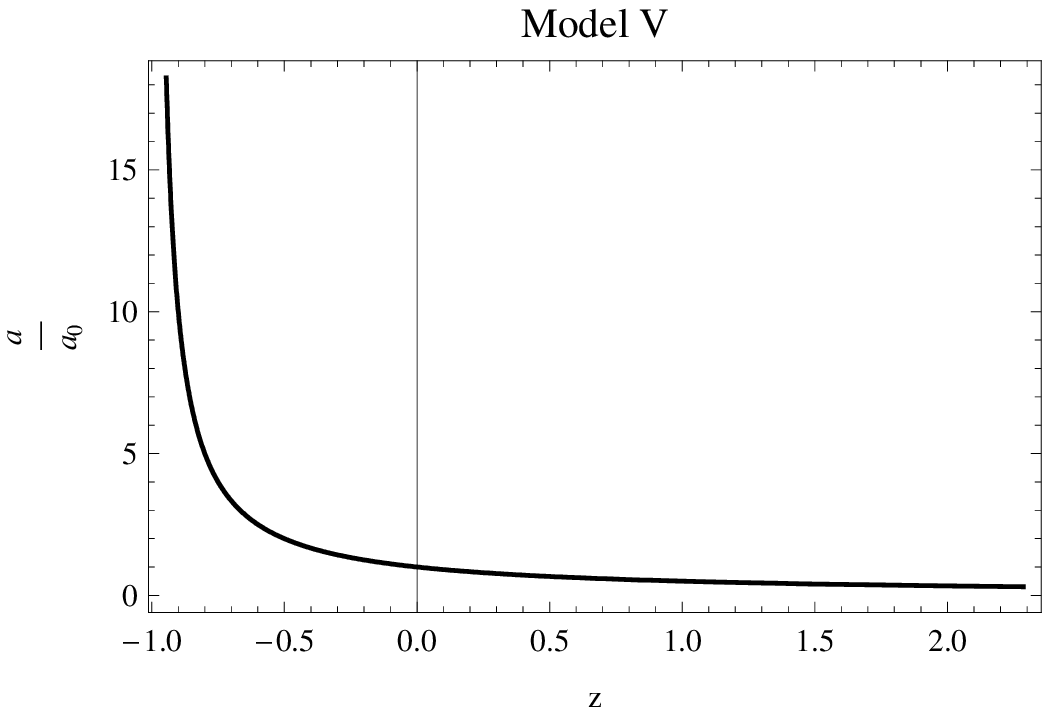}} \hspace{.2cm}
\subfigure[\label{fig5b} ]{ \includegraphics[width=.39\textwidth]%
{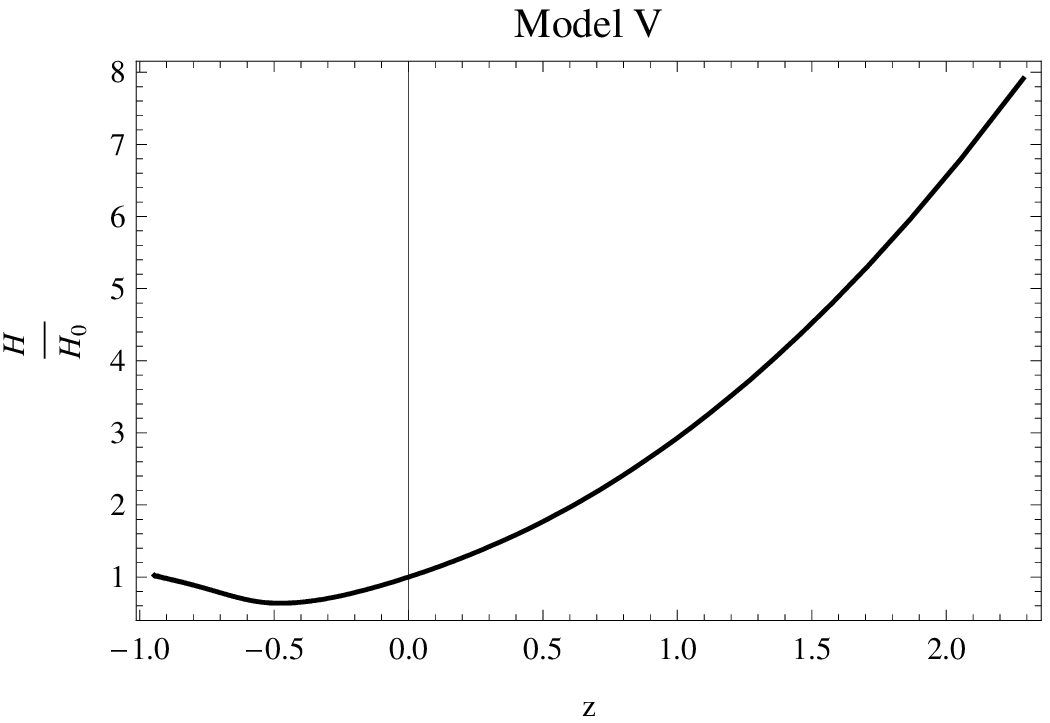}}
\end{minipage}
\\
\begin{minipage}[b]{1\textwidth}
\subfigure[\label{fig5c} ]{ \includegraphics[width=.39\textwidth]%
{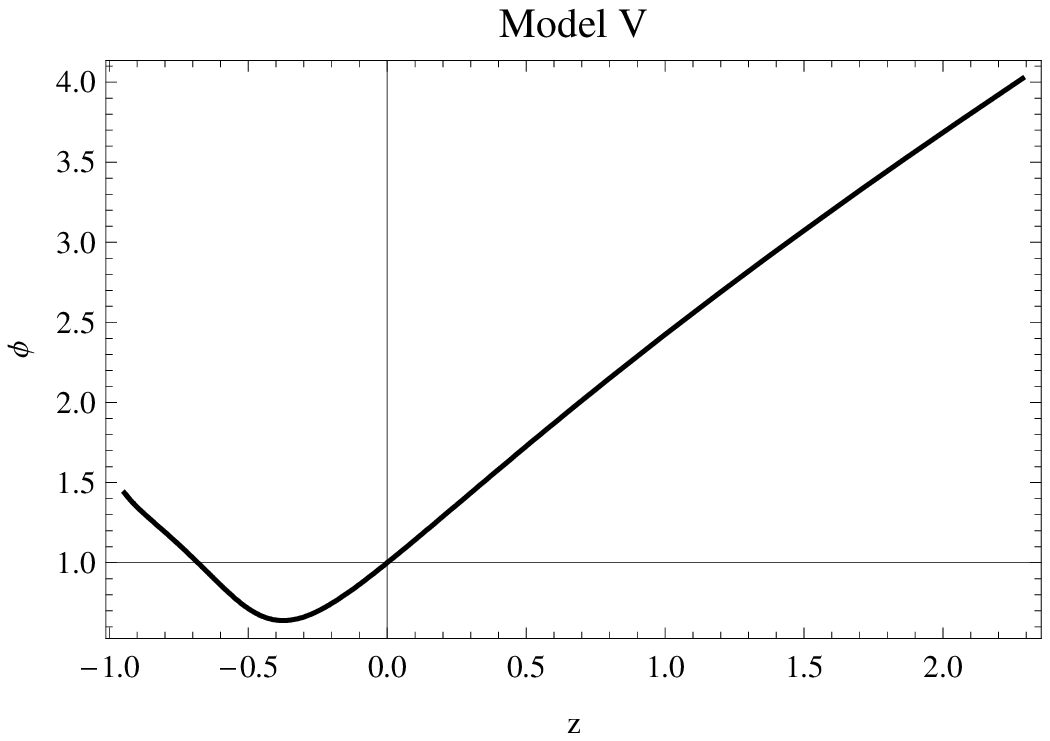}} \hspace{.2cm}
\subfigure[\label{fig5d} ]{ \includegraphics[width=.39\textwidth]%
{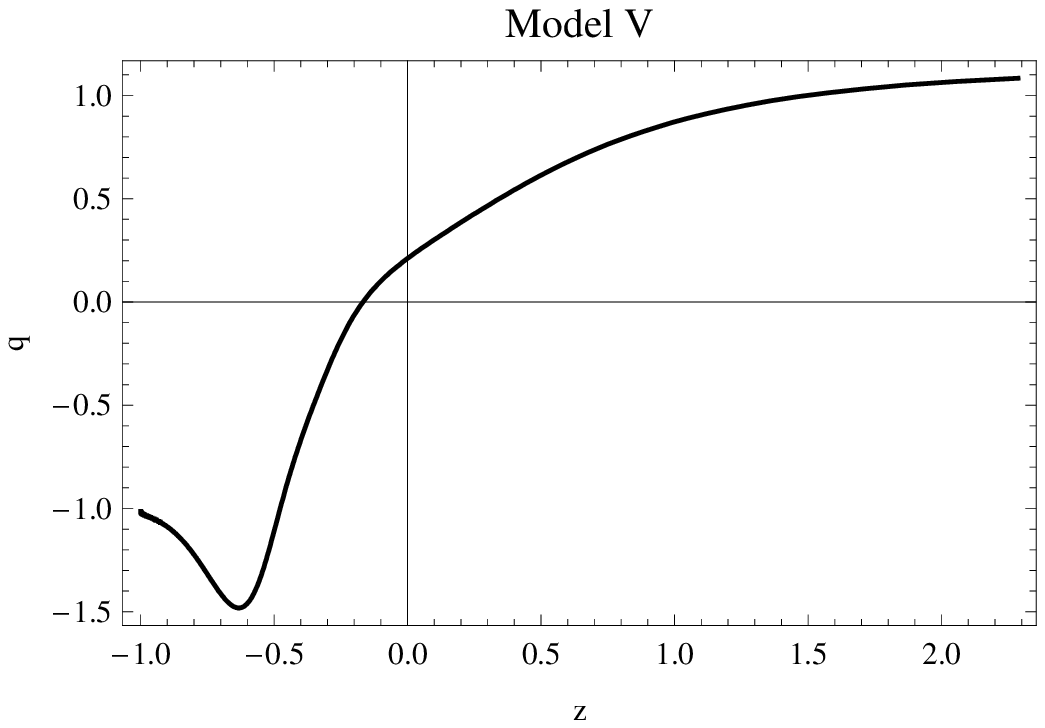}}
\end{minipage}
\\
\begin{minipage}[b]{1\textwidth}
\subfigure[\label{fig5e} ]{ \includegraphics[width=.39\textwidth]%
{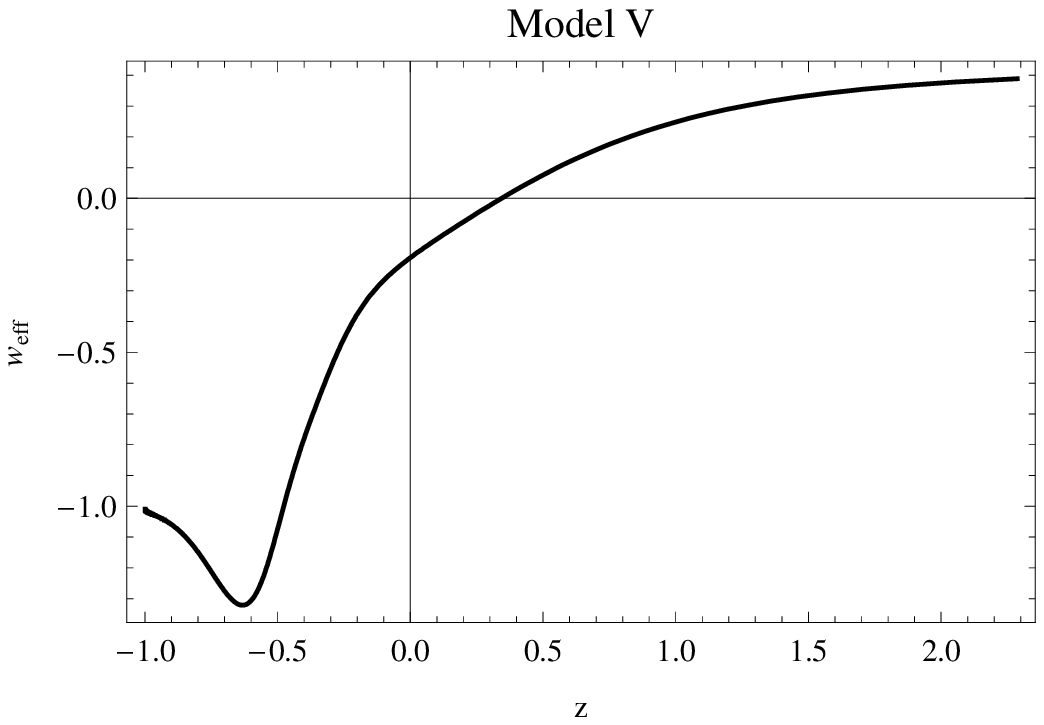}} \hspace{.8cm}
\subfigure[\label{fig5f} ]{ \includegraphics[width=.39\textwidth]%
{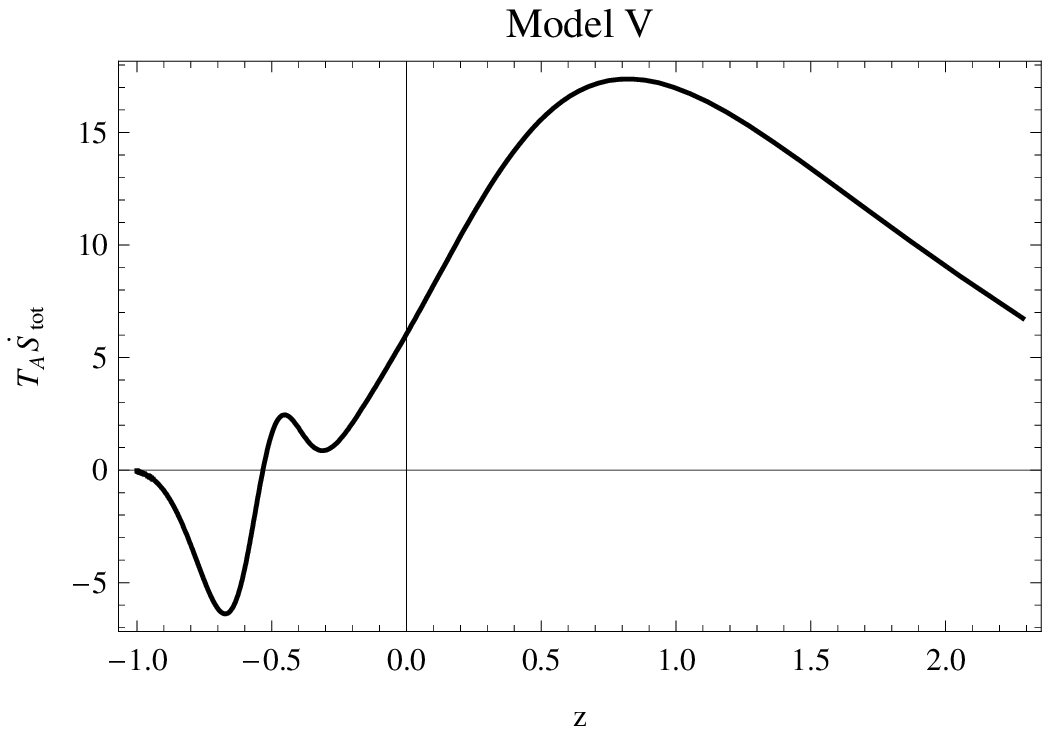}}
\end{minipage}
\caption{Same as Fig. \ref{ModelI} for model V (\ref{action5}).
Initial values are $a(1)=1$, $\dot{a}(1)=1$, $\phi(1)=1$ and
$\dot{\phi}(1)=-1.4$. Auxiliary parameters are: $\Omega_{\rm
m_0}=0.27$ \cite{2}$, \omega=1.2$, $f_0=-7$, $b=-0.4$
\cite{Farajollahi1} and $n=2$ \cite{Khoury}. Here $t_0=1/H_0$ and
$M^{n+4}=H_{0}^{2}$.} \label{ModelV}
\end{figure}
\end{document}